\documentclass[11pt]{article}
\usepackage{array,multirow}
\usepackage[utf8]{inputenc}
\usepackage[linesnumbered,ruled]{algorithm2e}
\usepackage{amsmath}
\usepackage{graphicx}
\usepackage{caption}
\usepackage{subcaption}
\usepackage{datetime}
\usepackage{longtable}
\usepackage{booktabs}
\usepackage{lscape}
\usepackage{hyperref}
\hypersetup{
    colorlinks = true,
    allcolors = blue
}
\usepackage{times}
\usepackage{float}
\restylefloat{table}
\usepackage{bm}
\usepackage[margin=1in]{geometry}

\usepackage[parfill]{parskip}
\usepackage{setspace}
\usepackage[title]{appendix}
\usepackage{dsfont}  % for double stroke fonts commonly used for indicators

\usepackage[dvipsnames,table]{xcolor}
\usepackage[table]{xcolor}
\usepackage{tikz}
\usetikzlibrary{backgrounds}

\usepackage{anyfontsize}

\usepackage{enotez} % for endnotes

\usepackage{amssymb}

\usepackage[inline]{enumitem}

\usepackage{amsmath}
\usepackage[round,authoryear]{natbib}
\usepackage{authblk}
\usepackage{tikz}
\usetikzlibrary{decorations.pathreplacing,positioning, arrows.meta}

\definecolor{mygray}{gray}{0.8}

\newcolumntype{L}[1]{>{\RaggedRight}p{#1}} % just a guess...

\definecolor{color1}{HTML}{BF2D2D} 
\definecolor{color2}{HTML}{CD5B5B} 
\definecolor{color3}{HTML}{DB8A8A} 
\definecolor{color4}{HTML}{E9B9B9} 
\definecolor{color5}{HTML}{F7E7E7} 
\definecolor{color6}{HTML}{E6EBE6} 
\definecolor{color7}{HTML}{B4C6B5} 
\definecolor{color8}{HTML}{83A084} 
\definecolor{color9}{HTML}{517953} 
\definecolor{color10}{HTML}{205423} 

\definecolor{textcolor1}{HTML}{737170}

\begin{document}
 \title{Modelling Interaction Duration in Relational Event Models}
%\title{DuREM: A Duration-Based Approach to Relational Event Modelling}

\author[1]{Rumana Lakdawala}
\author[1,2]{Roger Leenders}
\author[3]{Peter Ejbye-Ernst}
\author[1]{Joris Mulder}
\affil[1]{Tilburg University}
\affil[2]{Jheronimus Academy of Data Science}
\affil[3]{Netherlands Institute for the Study of Crime and Law Enforcement (NSCR)}
\date{}

 \maketitle
 \noindent\rule{\textwidth}{0.5pt}
\begin{abstract}
The study of relational events, which are interactions occurring between actors over time, has gained significant traction recently. Traditional relational event models typically focus on modelling the occurrence and sequence of events without considering their duration even though duration information is frequently available in empirical relational event data. We introduce a novel Duration Relational Event Model (DuREM) that incorporates the temporal duration of events into the analysis. The proposed model extends the existing framework by (i) allowing the inclusion of past event durations in the endogenous statistics to account for how the duration of past events affects the rate of future interactions, and (ii) extending the traditional relational event model by also modelling when events will end based on past event history and covariates. This is achieved by extending the risk set to include both ongoing events at risk of ending and idle dyads at risk of starting new events. The methodology is implemented in a new R package `durem'. Two case studies concerning team dynamics and inter-personal violence are presented to illustrate the applicability of the model.
 \end{abstract}
\noindent\rule{\textwidth}{0.5pt}

\section{Introduction}
The analysis of social interaction processes has helped to address significant empirical questions in areas such as development of conflicts \citep[e.g.,][]{Ejbye-Ernst2021,lakdawala2025bondscreatedequaldyadic,brandenberger_2020_InterdependenciesConflictDynamics,gravel_2023_RivalriesReputationRetaliation}, collaborative or organizational dynamics \citep{leenders_2016_OnceTimeUnderstanding,lerner_2019_REMDyadsRelational} and the coordination of emergency response efforts \citep{butts_2008_RelationalEventFramework}. To study these processes, several statistical methods have been developed, including dynamic models for social networks \citep{holland_1977_DynamicModelSocial,snijders_2010_IntroductionStochasticActorbased}, discrete temporal models \citep{hanneke_2010_DiscreteTemporalModels}, separable dynamic network models \citep{krivitsky_2014_SeparableModelDynamic}, and point process modelling approaches \citep{perry_2013_PointProcessModelling}. Among these, relational event models (REMs), introduced by \cite{butts_2008_RelationalEventFramework} and further developed by \cite{stadtfeld_2014_EventsSocialNetworks,stadtfeld_2017_DynamicNetworkActor,perry_2013_PointProcessModelling}, have become increasingly prominent. REMs are particularly valuable because they incorporate the history of past social interactions to examine how social behaviours evolve over time, offering a detailed understanding of temporal social processes.

REMs differ from other social network models by focusing on the temporal dynamics of interactions, rather than on static or aggregated representations of relationships. Traditional social network models often capture relationships as enduring ties or connections between actors and analyse patterns like density, clustering, or centrality. These models are useful for understanding the broader architecture of network structures but typically do not account for the timing or evolutionary nature of the interactions. In REMs, the time to the next relational event between two actors is directly modelled, thereby allowing researchers to study the impact of key predictors on the social interaction rates across actors over time. By incorporating both temporal and structural dependencies, REMs offer a more dynamic view of how ties evolve and change.

This temporal focus makes REMs particularly useful for studying phenomena where the order and timing of events are critical. For instance, understanding how and how fast information spreads through a network, how conflicts escalate, or how collaboration patterns emerge requires a model that captures not just who interacts with whom, but when and how often these interactions occur. REMs also allow for the inclusion of endogenous network factors such as reciprocity, transitivity, or inertia providing insights into how prior interactions shape future behaviours. These capabilities make REMs a powerful and flexible tool for examining a wide range of social interaction mechanisms in understanding network dynamics.

Standard REMs were originally developed for analysing instantaneous events between individuals observed at a single point in time, such as radio transmissions \citep[e.g.,][]{butts_2008_RelationalEventFramework}. However, in practice, many interactions have a duration attached to it. For instance the length of a phone call, the amount of time spent in conflict, the duration of a collaboration, and the length of face-to-face communication. As REMs aim to understand interaction dynamics between actors over time, modelling the duration of relational events is as important as modelling the time between events. If a dyad has frequent interaction or its interactions tend to have a very long duration, both imply that the actors in the dyad are `strongly connected' or have a strong relation with each other. Longer durations may indicate deeper engagement, greater investment, or more complications in a conflict. Alternatively, brief interactions may reflect fleeting exchanges, information exchange, or procedural communication. By incorporating duration, we capture a richer, more nuanced description of relational event data, which can reveal richer patterns of behaviour, relational dynamics, and the underlying structure of social networks. This perspective is especially relevant in contexts where the timing and length of interactions shape outcomes, such as negotiations, collaborations, or emergency crisis management. 

The influence of event duration has not been explored much however. One study explored how people allocate time to various activities within a day \citep{Marcum2015}, and another examined group entry and exit dynamics \citep{hoffman_2020_ModelDynamicsFacetoface}. However, little attention has been given to how the duration of past events themselves influences interaction rates or affect relational dynamics. This gap leaves open questions about the relationship between event duration and the dynamics of social interactions.

In this article we propose a duration relational event modelling framework (DuREM) for jointly modelling the duration of events and event occurrence. The proposed model extends the existing REM framework (i) by incorporating the duration of past events in the endogenous statistics allowing one to study how the duration of past events affects the interaction rates between dyads and (ii) by jointly modelling both the start and end of events based on the past event history and covariates. The first extension is achieved by  modelling the weight of past events according to their duration using an unknown parameter which is learned from the data. The second extension introduces a joint model for the rates at which events start and end. This is achieved by extending the risk set to include ongoing events that are at risk of ending, as well as idle dyads that are at risk of starting new events. 

The inclusion of duration-weighted endogenous statistics is comparable with REM extensions regarding memory effects where the weight of past events depends on the elapsed time since a past event was observed \citep[e.g.,][]{brandes_2009_NetworksEvolvingStep,arena_2023_HowFastWe,arena_2022_BayesianSemiParametricApproach}. Using a weighting scheme, analogous to the memory effect, the model can quantify the degree to which the duration of past events impacts interaction patterns.

The remainder of the paper is organized as follows. In Section \ref{sec:joint} we first introduce the standard Relational Event Model (REM) followed by introduction to the proposed Duration Relational Event Model (DuREM). We present two empirical applications of DuREM in Section \ref{sec:applications}: a case study on research team dynamics in Section \ref{sec:case1} followed by a case study on the dynamics of interpersonal violence in Section \ref{sec:case2}. The article ends with a brief discussion in Section \ref{sec:discuss} on the model limitations, connection to related works and future opportunities for research.

\section{Joint modelling the duration of events and event occurrence} \label{sec:joint}
In this section, we first present the standard relational event model and notations (Section \ref{sec:rem}). Next, we present a novel modelling framework for incorporating duration into the relational event model that allows for modelling of how duration influences the occurrence of future events as well as how past event history affects the duration of future events (Section \ref{sec:durem}).

\subsection{Standard Relational Event Model (REM)}
\label{sec:rem}
In a standard relational event model, we consider $ \mathbf{\mathcal{E}} = \{ o_1, o_2 \dots o_M \}$, a sequence of $M$ events observed in a continuous time interval $[0,t_M]$,  where $o_m = \{ i_m, j_m, t_m\}$ is a tuple of the sender, receiver, and time of the $m$'th observed event in the sequence $ \mathbf{\mathcal{E}}$. The riskset $\boldsymbol{\mathcal{R}}$ is the set of all events that can possibly be involved in an event. The composition of the riskset depends on the research setting. Often the riskset\footnote{We focus on a directed riskset here but undirected events can also be analysed with this method, as we will show in case study I.}  is defined as the set of all possible sender-receiver pairs  $\boldsymbol{\mathcal{R}}: (i,j) \subseteq \{\mathbf{A} \times \mathbf{A}, i\neq j \}$ among actors belonging to set $\mathbf{A}$.

The rate of events for a dyad $d = (i,j)$ at time $t$ is specified as a log-linear function of the predictors (more commonly referred to as ``statistics") and the corresponding parameters as follows:
\begin{equation}
    \lambda_{d}(t \ | \ \boldsymbol{\beta}, \mathbf{x}_{d}(\mathbf{\mathcal{E}}_t)) = 
     \exp\{  \boldsymbol{\beta}^{\mathsf{T}} \ \mathbf{x}_{d}(\mathbf{\mathcal{E}}_t) \} 
\end{equation}
where $\mathbf{x}_d(\mathbf{\mathcal{E}}_t)$ corresponds to the vector of statistics having length $P$ for dyad $d$ computed from the event history $\mathcal{E}_t$ observed until time $t$. The vector of statistics may consist of endogenous and/or exogenous statistics. Endogenous statistics summarise the past interactions between the dyads in the riskset  whereas the exogenous statistics capture the effect of nodal or dyadic attributes (or other ``external" characteristics) on the event rate. The coefficients $\boldsymbol{\beta}$ are the corresponding parameters that quantify the relative importance of each statistic on the event rate.

The likelihood of observing the event sequence $\mathbf{\mathcal{E}}_M$ is as follows:
\begin{equation}
    p(\mathbf{\mathcal{E}}_M | \ \boldsymbol{\beta}) = \prod_{m=1}^{M} p(o_m | \ \boldsymbol{\beta},\ \mathbf{X}(\mathbf{\mathcal{E}}_{t_{m-1}})) \\
\end{equation}
where $\mathbf{X}(\mathbf{\mathcal{E}}_t)$ contains all the statistics for all dyads in the riskset on the event sequence until time $t$. Furthermore, the probability of observing an event involving dyad $d_m$ at time $t_m$ is modelled by a multinomial distribution over the dyadic rates and the inter-event time between the $m$-th and $m-1$-th event is modelled by an exponential distribution:
\begin{equation}
    p(o_m | \ \boldsymbol{\beta},\ \mathbf{X}(\mathbf{\mathcal{E}}_{t_{m-1}})) \; = \; p(d_m | \ \boldsymbol{\beta},\ \mathbf{X}(\mathbf{\mathcal{E}}_{t_{m-1}})) \cdot p(t_m - t_{m-1} | \ \boldsymbol{\beta},\ \mathbf{X}(\mathbf{\mathcal{E}}_{t_{m-1}}))
\end{equation}
where
\begin{equation}
\begin{aligned}
    p(d_m | \ \boldsymbol{\beta},\ \mathbf{X}(\mathbf{\mathcal{E}}_{t_{m-1}}) = \; & \; \frac{ \lambda_{d_m}(t_m \ | \ \boldsymbol{\beta}, \mathbf{x}_{d}(\mathbf{\mathcal{E}}_{t_{m-1}}))}{\sum \limits_{k \in \boldsymbol{\mathcal{R}}}  \lambda_{k}(t_m \ | \ \boldsymbol{\beta}, \mathbf{x}_{k}(\mathbf{\mathcal{E}}_{t_{m-1}})) } \\
    p(t_m - t_{m-1} | \ \boldsymbol{\beta},\ \mathbf{X}(\mathbf{\mathcal{E}}_{t_{m-1}}) = \; & \; \text{Exponential} \left(\sum \limits_{d \in \boldsymbol{\mathcal{R}}}  \lambda_{d}(t_m \ | \ \boldsymbol{\beta}, \mathbf{x}_{d}(\mathbf{\mathcal{E}}_{t_{m-1}})) \right).
\end{aligned}
\end{equation}

The relational event model is equivalent to a non-homogeneous multivariate Poisson counting process on the dyads $d=(i,j)$:
\begin{equation}
    \mathbf{N}(t) = ({N}_d(t) \ | \ d \in \boldsymbol{\mathcal{R}}),
\end{equation}
where each element ${N}_d(t)$ of $\mathbf{N}(t)$, indicates the cumulative count of events for the dyad $d$, i.e from $i$ to $j$ in the time interval $[0,t)$. %Moreover, we define $\Delta \mathbf{N}(t_1, t_2) = \mathbf{N}(t_2) - \mathbf{N}(t_1)$ for $t_2 >t_1$, where the $d$-th element, $\Delta \mathbf{N}(t_1, t_2)$, denotes the count of events that occurred for dyad $d$ in the interval $ [t_{1},t_2)$.

It is important to note that standard REMs do not directly model the time between consecutive interactions involving the particular actors. Instead, the focus is on the timing of sequential events across the entire network, considering each event relative to all other events. Therefore, the waiting time or interevent time modelled by REM reflects the interval until the next event occurs anywhere within the network, rather than the interval until the next event within a specific actor pair (dyad). %This property becomes particularly relevant when extending the REM framework to incorporate event duration, as modelling event duration explicitly introduces considerations about how ongoing interactions between specific actors influence the timing of subsequent events. The implications of this distinction will be elaborated further in the following section.

\subsection{Duration Relational Event Model (DuREM)}
\label{sec:durem}
We extend the REM by allowing the incorporation of event durations into the framework of relational event analysis. The extended framework will be referred to as the Durational Relational Event Model (DuREM). In a DuREM, each event is characterized not only by its start time $t^s_m$ but also by its end time $t^e_m$ as well. Events are therefore defined as $o_m = (i_m, j_m, t^s_m, t^e_m)$, where $i_m$ and $j_m$ are the sender and receiver of the event, $t^s_m$ is the event's start time, and $t^e_m$ is the event's end time. %The duration of an event is then given by $u_m = t^e_m - t^s_m$.

The risk set in a DuREM, $\boldsymbol{\mathcal{R}}(t)$, is a union of two disjoint composites that vary with time: $ \boldsymbol{\mathcal{R}}(t) = \boldsymbol{\mathcal{R}}^s(t) \cup \boldsymbol{\mathcal{R}}^e(t)$, where $\boldsymbol{\mathcal{R}}^s(t)$ contains all possible dyads that are risk to start an event at time $t$ and $\boldsymbol{\mathcal{R}}^e(t)$ consists of all active dyads that are currently involved in an event and are at risk to end their event at time $t$. %We can formalize whether dyad $d=(i,j)$ at time $t$ is involved in an ongoing event by an indicator variable $\mathcal{J}_{d}$, such that  $\mathcal{J}_{d}(t) = 1$ if it is and $0$ if it is not: 
We can formalize whether a dyad $d=(i,j)$ at any time $t$ is at risk to start or end an event by an indicator variable $\mathcal{J}_{d}(t)$, such that  $\mathcal{J}_{d}(t) = 0$ if it is at risk to start an event and $1$ if it currently involved in an ongoing event and is at risk to end its current event: 
%We can formalize whether dyad $d=(i,j)$ at time $t$ is at risk to start or end an event by an indicator variable $\mathcal{J}_{d}(t)$, such that  $\mathcal{J}_{d}(t) = 0$ if it is at risk to start an event and $1$ if it is at risk to end its current event: 
\begin{equation}
    \mathcal{J}_{d}(t) = \begin{cases}       
        0 &  \text{if} \ d \in \boldsymbol{\mathcal{R}}^s(t). \\
        1 & \text{if} \  d \in \boldsymbol{\mathcal{R}}^e(t) \\
    \end{cases}
\end{equation}

The definition of the riskset and rate parameters of the DuREM is very related to the standard REM with typed events \citep{butts_2008_RelationalEventFramework}. In that model, events can have different types (e.g., face-to-face interaction or digital interaction) and both event types are jointly at risk. The DuREM, in a way, also consists of two event types, namely an end event (if two actors end their event they were in) or a start event (if two actors start an event). The difference is that in the DuREM each dyad can only be in either the riskset to start an event or in the riskset to end an event, while in the REM with typed events each dyad can (in principle) be part of both risksets. 

We model the dynamics of event initiation and ending using two non-homogeneous matrix-valued counting processes: $ \mathbf{N}^{\text{s}}(t) = (\mathbf{N}^{s}_d(t) \; |  \;  d \in \boldsymbol{\mathcal{R}}^s)$ for starting events and  $\mathbf{N}^{\text{e}}(t) = (\mathbf{N}^{e}_d(t) \; |  \;  d \in \boldsymbol{\mathcal{R}}^e)$ for ending the events.
These processes are characterised by their respective intensities $\lambda^s_d(t)$ and $\lambda^e_d(t)$, which are functions of the event history and covariates. The intensity for starting an event is given by: 
\begin{equation}
    \lambda^s_{d}(t \ | \ \boldsymbol{\beta}^s, \mathbf{x}^s_{d}(\mathbf{\mathcal{E}}_t)) = \begin{cases} 
     \exp\{  \boldsymbol{\beta}^{s\mathsf{T}} \ \mathbf{x}^s_{d}(\mathbf{\mathcal{E}}_t) \} & \  \text{if} \ \mathcal{J}_{d}(t) = 0 \\
    0 & \ \text{if} \ \mathcal{J}_{d}(t) = 1
    \end{cases}
\end{equation}
and the intensity for ending an event is:
\begin{equation}
    \lambda^e_{d}(t \ | \ \boldsymbol{\beta}^e, \mathbf{x}^e_{d}(\mathbf{\mathcal{E}}_t)) = \begin{cases} 
    0 & \ \text{if} \  \mathcal{J}_{d}(t) = 0\\
    \exp\{  \boldsymbol{\beta}^{e\mathsf{T}} \ \mathbf{x}^e_{d}(\mathbf{\mathcal{E}}_t) \} & \ \text{if} \ \mathcal{J}_{d}(t) = 1
    \end{cases}
\end{equation}
where we denote the statistics by $\mathbf{x}^s_{d}(\mathbf{\mathcal{E}}_t)$ for the start model and $\mathbf{x}^e_{d}(\mathbf{\mathcal{E}}_t)$ for the end model. Similar to the standard REM, these statistics summarise the event history and relevant covariates that influence the propensity of dyads to initiate or terminate events. The corresponding parameters $\boldsymbol{\beta}^s$ and $\boldsymbol{\beta}^e$ quantify the effects of these covariates on corresponding the event rates.

\subsubsection{Directed and Undirected Model Specification for Start and End Rate Models}

The DuREM can be used for either directed or undirected events, leading to four possible model configurations for $\mathcal{R}^s$ and $\mathcal{R}^e$. The choice depends on whether event initiation and termination are symmetrical or asymmetrical, which varies based on the research and empirical context.

When both the start and end of an event are undirected (UU), either participant in the dyad can initiate or terminate the interaction. This is useful when the role of the initiator is irrelevant to the analysis, such as in proximity sensor data that records when two individuals come into contact but does not indicate who initiated or ended the interaction. It is also applicable when only the interaction and its duration are relevant, rather than which actor initiated or ended an event. When both the start and end of an event are directed (DD), initiation and termination are intentional actions taken by specific actors. An example is a scheduled meeting, where one person arranges and starts it (directed start), and the meeting concludes when one of the participants decides to it (directed end).

When the start of an event is directed but the end is undirected (DU), a specific actor initiates the event, but once it begins, both participants engage equally, and the interaction ends naturally. An example is a physical fight; one person starts the altercation (directed start), but once engaged, both contribute equally, and it typically ends due to external intervention or mutual agreement (undirected end). In cases where the start is undirected but the end is directed (UD), events begin symmetrically but are ended by a specific individual. This applies to situations where interactions are assigned externally but later terminated by one of the participants. A relevant example is a friendship-matching app where users are algorithmically paired (undirected start), but either party can independently decide to end the interaction (directed end).  

Together, these four configurations for the start and risksets, i.e. $\boldsymbol{\mathcal{R}}^s$ and $\boldsymbol{\mathcal{R}}^e$ provide flexibility in modelling different types of duration relational event dynamics.

\subsubsection{Capturing the effect of duration on start and end event rate}
The inclusion of weights for start and end events can enable the model to explicitly incorporate the duration of past events into the rate of future events. We propose a weighing scheme that aims to capture the dependencies between event occurrence and duration, where longer durations may either enhance or inhibit the probability of future interactions depending on the context. %The weight for each observed event $o_m$ for dyad $d=(i,j)$ are defined as:
We define the weight of a single past event $o_m$ between actors $i$ and $j$, occurring at times $t^s_m$ (start) and $t^e_m$ (end), based on its duration:
\begin{equation}
w_m^{s, \text{duration}}(t, i, j) = (t^e_m - t^s_m)^{\psi^s}
\label{eqn:w_duration_s}
\end{equation}
\begin{equation}
w_m^{e, \text{duration}}(t, i, j) = (t^e_m - t^s_m)^{\psi^e}.
\label{eqn:w_duration_e}
\end{equation}

% NEW FORMULA
% \begin{equation}
% \begin{aligned}
% w_{m}^{s,\mathrm{duration}}(t,i,j)
% &=
% \exp\!\left(
% \psi_s  \left(t_m^e - t_m^s\right)
% \right),
% \\
% w_{m}^{e,\mathrm{duration}}(t,i,j)
% &=
% \exp\!\left(
% \psi_e  \left(t_m^e - t_m^s\right)
% \right).
% \end{aligned}
% \end{equation}

% \begin{equation}
%     w^s(t^s,i,j) = \sum \limits_{o_m:(i_m,j_m) = (i,j), t_m<t} (t^o_m-t^s_m)^{\psi^s} 
%     \label{eqn:w_s}
% \end{equation}
%  \begin{equation}    
%     %\quad 
%     w^e(t^e,i,j) = \sum \limits_{o_m:(i_m,j_m) = (i,j), t_m<t} (t^o_m-t^s_m)^{\psi^e}
%     \label{eqn:w_e}
% \end{equation}
Here, $\psi^s$ and $\psi^e$ are parameters that determine the impact of event duration on the rate of starting or ending subsequent events. These parameters can be learnt or estimated from data. When $\psi^s > 0$, past events having a longer duration have a higher contribution to the weight of subsequent interactions between $i$ and $j$. When $\psi^s = 0$, the weight becomes independent of duration; therefore, the rate of starting events is not influenced by how long the past interactions lasted. When $\psi^s < 0$, events having a longer duration decrease the weight of starting subsequent interactions. Similarly, when $\psi^e > 0$, longer duration events contribute to a higher weight, when $\psi^e = 0$, the duration of past events does not influence the end rate of subsequent interactions, and when $\psi^e < 0$: longer duration events decrease the weight of ending events. 

The weight is used primarily in the computation of the endogenous statistics in $\mathbf{x}^s_{d}(\mathbf{\mathcal{E}}_t))$ for the start model and $\mathbf{x}^e_{d}(\mathbf{\mathcal{E}}_t))$ for the end model. For example, the standard REM the statistic \textit{inertia}, which is computed as the count of all the past events between actors $i$ and $j$: 

\begin{equation*}
    \text{inertia}(t_m, i, j) = \sum_{m=1}^{M} \mathbb{I}(o_m : (i_m, j_m) = (i, j)), 
\end{equation*}

where $\mathbb{I}$ indicates whether the dyad $(i, j)$ were observed in the event $o_m$ at time $t_m$. In the DuREM framework, inertia is modified for the start and end models by computing the weighted sum of all past events $o_m$ that occurred between the actors $i$ and $j$: 

\begin{equation}
\text{inertia}^{s}(t, i, j) = \sum \limits_{o_m:(i_m, j_m) = (i, j), t^s_m<t} w_m^{s, \text{duration}}(t, i, j)
\label{eqn:w_aggregate_duration_s}
\end{equation}

\begin{equation}
\text{inertia}^{e}(t, i, j) = \sum \limits_{o_m:(i_m, j_m) = (i, j), t^e_m<t} w_m^{e, \text{duration}}(t, i, j)
\label{eqn:w_aggregate_duration_e}
\end{equation}

In addition to capturing the influence of event duration, the model can also account for the influence of memory. Actors’ forgetfulness may cause the impact of past events on future interactions to decrease over time, a phenomenon often referred to as the ``memory effect" \citep{brandes_2009_NetworksEvolvingStep,arena_2022_BayesianSemiParametricApproach,arena_2023_HowFastWe,meijerink-bosman_2022_DynamicRelationalEvent}. This memory effect can be incorporated into the weighting scheme by an exponential decay function that reduces the influence of past events as time elapses. The resulting memory-specific weights for the start and end models are defined as:

% \begin{equation}
%     w^s(t^s, i, j) = \left[ \sum \limits_{o_m:(i_m, j_m) = (i, j), t_m < t} (t^o_m - t^s_m)^{\psi^s} \right] \cdot \exp \left[ - (t - t^o_m) \frac{\ln(2)}{\tau} \right] \frac{\ln(2)}{\tau}
% \end{equation}

% \begin{equation}
%     w^e(t^e, i, j) = \left[ \sum \limits_{o_m:(i_m, j_m) = (i, j), t_m < t} (t^o_m - t^s_m)^{\psi^e} \right] \cdot \exp \left[ - (t - t^o_m) \frac{\ln(2)}{\tau} \right] \frac{\ln(2)}{\tau}
% \end{equation}
\begin{equation}
w_m^{s, \text{memory}}(t, i, j) = \exp\left[-(t - t^s_m)\frac{\ln(2)}{\tau}\right] \frac{\ln(2)}{\tau}
\label{eqn:w_memory_s}
\end{equation}

\begin{equation}
w_m^{e, \text{memory}}(t, i, j) = \exp\left[-(t - t^e_m)\frac{\ln(2)}{\tau}\right] \frac{\ln(2)}{\tau}
\label{eqn:w_memory_e}
\end{equation}

%The former terms captures the influence of the event duration, while the exponential decay term models the memory effect. 
The parameter $\tau$ represents the half-life of the memory effect (ie. the time after which the weight of an event is halved). The factor $\frac{\ln(2)}{\tau}$ ensures that the memory effect is scaled appropriately, for the the aggregate influence per unit time \citep{lerner_2013_ModelingFrequencyType,arena_2022_BayesianSemiParametricApproach}.

Therefore the total weights for a past event incorporating both duration and memory effects are given by:

\begin{equation}
w_m^s(t, i, j) = w_m^{s, \text{duration}}(t, i, j) \cdot w_m^{s, \text{memory}}(t, i, j)
\label{eqn:w_single_s}
\end{equation}

\begin{equation}
w_m^e(t, i, j) = w_m^{e, \text{duration}}(t, i, j) \cdot w_m^{e, \text{memory}}(t, i, j)
\label{eqn:w_single_e}
\end{equation}

Thus, inertia in the DuREM framework, accounting for both event duration and memory effects, is defined for start and end rates as:

\begin{equation}
\text{inertia}^s(t, i, j) = \sum \limits_{o_m:(i_m, j_m) = (i, j), t^s_m<t} w^s(t, i, j)
\label{eqn:inertia_weighted_s}
\end{equation}

\begin{equation}
\text{inertia}^e(t, i, j) = \sum \limits_{o_m:(i_m, j_m) = (i, j), t^e_m<t}  w^e(t, i, j)
\label{eqn:inertia_weighted_e}
\end{equation}

The modified DuREM formulations for additional endogenous statistics, along with their interpretations, are provided in Appendix \ref{sec:stats}. Table \ref{tab:durem-stats} presents an overview of the available endogenous statistics for directed relational events, while Table \ref{tab:durem-undirected-stats} provides an overview for undirected events.

\subsubsection{Model for the start and end rate}
As the standard REM aims to explain when the next event will be observed and which actors will be part of the event, without directly modelling the time between events for individual pairs of actors, DuREM similarly seeks to explain the timing of the next event. This event can either be the start of a new event (for dyads in $\boldsymbol{\mathcal{R}}^s$ that are currently at risk to start a relational event) or the termination of an ongoing interaction (for dyads in $\boldsymbol{\mathcal{R}}^e$ that are currently involved in an event and are at risk to end it). Therefore, the parameters of the DuREM should be interpreted in a similar manner as those in the standard REM.

The likelihood function for the combined start and end ($2 \times M$) events  incorporates these intensities and weights and is expressed as:

% \begin{equation}
%     p(\mathbf{\mathcal{E}}_{M} \mid \boldsymbol{\beta}^s, \boldsymbol{\beta}^e, \psi^s, \psi^e) = 
%     \prod_{m=1}^{2M} 
%     p(d_m, \mathcal{J}_{d_m}(t_m) \mid \boldsymbol{\beta}^s, \boldsymbol{\beta}^e, \psi^s, \psi^e, \mathbf{X}(\mathbf{\mathcal{E}}_{t_{m-1}})) \cdot 
%     p(t_m - t_{m-1} \mid \boldsymbol{\beta}^s, \boldsymbol{\beta}^e, \psi^s, \psi^e, \mathbf{X}(\mathbf{\mathcal{E}}_{t_{m-1}}))
%     \label{eqn:lik}
% \end{equation}
\begin{equation}
    p(\mathbf{\mathcal{E}}_{M} \mid \boldsymbol{\beta}^s, \boldsymbol{\beta}^e, \psi^s, \psi^e) 
    \begin{aligned}[t]
        & \\
        &\hspace{-70pt} = \prod_{m=1}^{2M} 
        p(d_m, \mathcal{J}_{d_m}(t_m) \mid \boldsymbol{\beta}^s, \boldsymbol{\beta}^e, \psi^s, \psi^e, \mathbf{X}(\mathbf{\mathcal{E}}_{t_{m-1}})) \cdot p(t_m - t_{m-1} \mid \boldsymbol{\beta}^s, \boldsymbol{\beta}^e, \psi^s, \psi^e, \mathbf{X}(\mathbf{\mathcal{E}}_{t_{m-1}}))
    \end{aligned}
    \label{eqn:lik}
\end{equation}

with

\begin{equation}
    p(d_m, \mathcal{J}_{d_m}(t_m) \mid \boldsymbol{\beta}^s, \boldsymbol{\beta}^e, \psi^s, \psi^e, \mathbf{X}(\mathbf{\mathcal{E}}_{t_{m-1}}))
    \begin{aligned}[t]
    & \\
    & \hspace{-70pt} = \frac{\bigl(\lambda_{d_m}^s(t_m \mid \boldsymbol{\beta}^s, \psi^s, \mathbf{x}^s_{d}(\mathbf{\mathcal{E}}_{t_{m-1}}))\bigr)^{1 - \mathcal{J}_{d_m}(t_m)} \cdot \bigl(\lambda_{d_m}^e(t_m \mid \boldsymbol{\beta}^e, \psi^e, \mathbf{x}^e_{d}(\mathbf{\mathcal{E}}_{t_{m-1}}))\bigr)^{\mathcal{J}_{d_m}(t_m)}}{\sum \limits_{k \in \boldsymbol{\mathcal{R}}^s(t_m)} \lambda^s_k(t_m \ | \ \boldsymbol{\beta}^s, \psi^s, \mathbf{x}^s_{d}(\mathbf{\mathcal{E}}_{t_{m-1}})) + \sum \limits_{k \in \boldsymbol{\mathcal{R}}^e(t_m)} \lambda^e_k(t_m \ | \ \boldsymbol{\beta}^e, \psi^e, \mathbf{x}^e_{d}(\mathbf{\mathcal{E}}_{t_{m-1}}))}
    \end{aligned}
\end{equation}

where the $p(d_m, \mathcal{J}_{d_m}(t_m) \mid \boldsymbol{\beta}^s, \boldsymbol{\beta}^e, \psi^s, \psi^e, \mathbf{X}(\mathbf{\mathcal{E}}_{t_{m-1}}))$ represents the probability of observing the specific dyad \(d_m\) either starting ( if $\mathcal{J}_{d_m}(t_m) = 0$) or ending an event (if $\mathcal{J}_{d_m}(t_m) = 1$) at time $t_m$. The inter-event time between the events as exponentially distributed as follows:

\begin{equation}
\begin{aligned}
    p(t_m - t_{m-1} \mid  \boldsymbol{\beta}^s, \boldsymbol{\beta}^e, \psi^s, \psi^e, \mathbf{X}(\mathbf{\mathcal{E}}_{t_{m-1}})) \\ 
    & \hspace{-15em} = \text{Exponential}\left(\sum_{d \in \boldsymbol{\mathcal{R}}^s(t_m)} \lambda^s_d(t_m \ | \ \boldsymbol{\beta}^s, \psi^s, \mathbf{x}^s_{d}(\mathbf{\mathcal{E}}_{t_{m-1}})) + \sum_{d \in \boldsymbol{\mathcal{R}}^e(t_m)} \lambda^e_d(t_m \ | \ \boldsymbol{\beta}^e, \psi^e, \mathbf{x}^e_{d}(\mathbf{\mathcal{E}}_{t_{m-1}}) )\right)
\end{aligned}
\end{equation}

The rate parameter for this exponential distribution is the cumulative rate of all possible start and end events in $\boldsymbol{\mathcal{R}}^s(t) \cup \boldsymbol{\mathcal{R}}^e(t)$ .

This setting is inspired from the competing risks models \citep{crowder_2001_ClassicalCompetingRisks} from survival analysis where events (such as death) can be caused by one of several competing reasons. In survival analysis, competing risks models are utilized to understand how different causes can influence the timing and likelihood of an event. In our case, the DuREM extends this principle to relational events, modelling the simultaneous competition between all start-type and end-type events to determine which will be the next event observed. Each dyad in the riskset has the potential to either initiate a new event or conclude an ongoing one, with the probabilities of these occurrences being dynamically shaped by the event history and associated covariates. This competition between start and end events allows the model to capture the complex processes that drive event initiation, event persistence, and termination within relational networks.

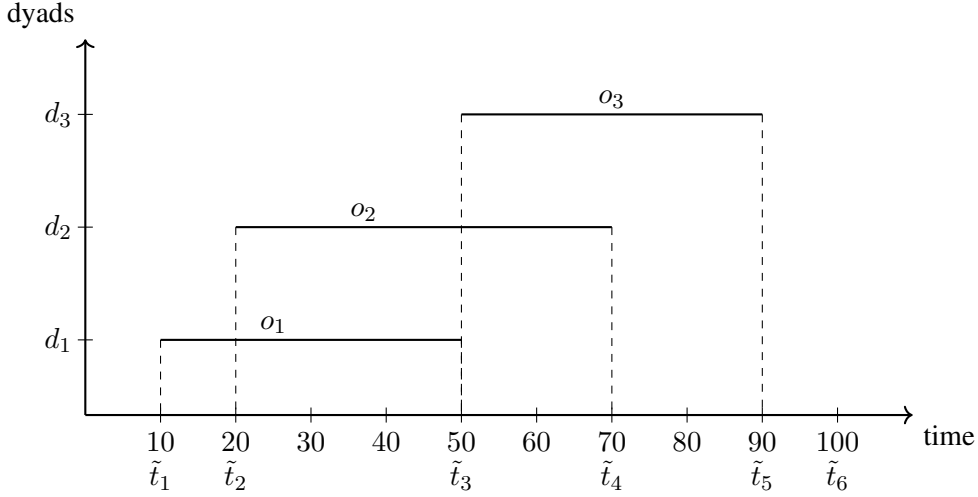
\begin{figure}[t]
    \centering
    \begin{tikzpicture}
    % Draw the axes
    \draw[thick,->] (0,0) -- (11,0) node[anchor=north west] {time};
    \draw[thick,->] (0,0) -- (0,5) node[anchor=south east] {dyads};

    % Draw ticks and labels for dyad d_1, d_2, d_3
    \foreach \y/\label in {1/d_1, 2.5/d_2, 4/d_3} {
        \draw (-0.1,\y) -- (0.1,\y) node[left=4pt] {$\label$};
    }

    % Draw line segments for each dyad and add labels
    % d_1 from (1,1) to (4,1)
    \draw[thick] (1,1) -- (5,1);
    \node at (2.5, 1.2) {$o_1$}; % Label for the first segment
    \draw[dashed] (1,1) -- (1,0);
    \draw[dashed] (5,1) -- (5,0);
    
    % d_2 from (2,2.5) to (6,2.5)
    \draw[thick] (2,2.5) -- (7,2.5);
    \node at (3.7, 2.7) {$o_2$}; % Label for the second segment
    \draw[dashed] (2,2.5) -- (2,0);
    \draw[dashed] (7,2.5) -- (7,0);
    
    % d_3 from (3,4) to (9,4)
    \draw[thick] (5,4) -- (9,4);
    \node at (7, 4.2) {$o_3$}; % Label for the third segment
    \draw[dashed] (5,4) -- (5,0);
    \draw[dashed] (9,4) -- (9,0);

    % Add x-axis numbers
    \foreach \x in {1,2,...,8,9,10}
    \draw (\x,0.1) -- (\x,-0.1) node[below] {$\x0$};

    % Add x-axis special labels below numbers
  \foreach \x/\label in {1/\tilde{t}_1, 2/\tilde{t}_2, 5/\tilde{t}_3, 7/\tilde{t}_4, 9/\tilde{t}_5, 10/\tilde{t}_6} {
    \node at (\x, -0.8) {$\label$};
}
\end{tikzpicture}

    \caption{Representation of sequence of start and end relational events for dyads $d_1$, $d_2$, and $d_3$.}
    \label{fig:durem-rs}
\end{figure}

By modelling these competing processes, the DuREM can address questions such as how the duration of ongoing events affects the likelihood of new events starting or existing ones ending and how the accumulation of past events influences the dynamics of future events. For instance, a longer ongoing event might either increase or decrease the rate of a new interaction involving the same dyad, depending on the context and the covariates in $\mathbf{x}^s_{d}(\mathbf{\mathcal{E}}_t)$ and $\mathbf{x}^e_{d}(\mathbf{\mathcal{E}}_t)$. %This flexibility in incorporating both event durations and past event history makes the DuREM particularly suitable for analysing dynamic relational histories where events may overlap or begin simultaneously.

Figure \ref{fig:durem-rs} illustrates the competing risks framework within the context of the DuREM model by representing the temporal dynamics of directed start and end events for dyads $d_1$, $d_2$, and $d_3$. Each horizontal line corresponds to an event, starting and ending at specific time points, marked by $\tilde{t}_1, \tilde{t}_2, \dots, \tilde{t}_6$. The dashed vertical lines denote the duration points for events, for when a dyad starts or ends an event.

\begin{table}[h]
    \centering
    \begin{minipage}[b]{0.45\textwidth}
        \begin{table}[H]
            \centering
            \begin{tabular}{c|c|c}
                dyad & start time $t^s$ & end time $t^e$\\
                \hline
                $d_1$ & 10 & 50 \\
                $d_2$ & 20 & 70 \\
                $d_3$ & 50 & 90 \\ 
            \end{tabular}
            \caption{Sequence of start and end event observations for a duration relational event model.}
            \label{tab:interval-table}
        \end{table}
    \end{minipage}
    \hfill
    \begin{minipage}[b]{0.5\textwidth}
        \begin{table}[H]
            \centering
            \begin{tabular}{c|c|c}
                time & $\boldsymbol{\mathcal{R}}^s$ (at risk to start) & $\boldsymbol{\mathcal{R}}^e$ (at risk to end) \\
                \hline
                $\tilde{t}_1 = 10$ & $\{d_1,d_2,d_3\}$ & $\emptyset$ \\
                $\tilde{t}_2 = 20$ & $\{d_2,d_3\}$ & $\{d_1\}$ \\
                $\tilde{t}_3 = 50$ & $\{d_3\}$ & $\{d_1, d_2\}$ \\
                $\tilde{t}_4 = 70$ & $\{d_1\}$ & $\{d_2,d_3\}$ \\
                $\tilde{t}_5 = 90$ & $\{ d_1,d_2\}$ & $\{d_3\}$ \\
                $\tilde{t}_6 = 100$ & $\{d_1,d_2,d_3\}$ & $\emptyset$ \\
            \end{tabular}
           \caption{Corresponding riskset compositions for starting ($\boldsymbol{\mathcal{R}}^s$) and ending ($\boldsymbol{\mathcal{R}}^e$) events.}
            \label{tab:event-timeline}
        \end{table}
    \end{minipage}
\end{table}

Table \ref{tab:interval-table} represents how a typical duration relational event history data would be represented and Table \ref{tab:event-timeline} illustrates the construction of the dynamic risk-sets for DuREM. %For example, at time $t_3$, the risk set includes three potential events: $\boldsymbol{\mathcal{R}}(\tilde{t}_3)= \{\text{end}(d_1), \text{end}(d_2), \text{start}(d_3)\}$. The event involving $d_1$ has been ongoing since $\tilde{t}_1$ and is at risk of ending at $tilde{t}_3$. The event involving $d_2$, which began at $tilde{t}_2$, is at risk of ending in the future. A new event involving $d_3$ is at risk of starting at $\tilde{t}_3$. 
For example at time $\tilde{t}_2 = 20$, dyads $d_2$ and $d_3$ are at risk of starting an event, i.e., $\mathcal{R}^s(\tilde{t}_2) = \{d_2, d_3\}$. Meanwhile, dyad $d_1$, which initiated an event at $\tilde{t}_1 = 10$, is still ongoing and thus at risk of ending, so $\mathcal{R}^e(\tilde{t}_2) = \{d_1\}$. At time $\tilde{t}_3 = 50$, dyad $d_1$ is still ongoing, and dyad $d_2$, which began at $\tilde{t}_2 = 20$, is also ongoing. Therefore, $\mathcal{R}^e(\tilde{t}_3) = \{d_1, d_2\}$. Whereas, dyad $d_3$ is at risk of starting, $\mathcal{R}^s(\tilde{t}_3) = \{d_3\}$. This flexibility in modelling both event start and end makes the DuREM particularly suitable for analysing dynamic relational histories where events may overlap or begin simultaneously.

\subsection{Parameter Estimation Using the \texttt{durem} Package}
We consider a maximum likelihood estimation approach to fit the DuREM. The maximum likelihood estimates (MLEs) maximize the likelihood presented in Equation \ref{eqn:lik}. However, the time-varying risk set and the inclusion of event durations must be accounted for in the statistics computation. Conceptually, this entails processing the observed event sequence, updating the risk set after each event, and incorporating the weight factors $w^s$ and $e^e$ as specified in the model. These weights enable the model to capture the effect that the durations of past events have on future event dynamics. Once the model’s intensities $\lambda^s$ and $\lambda^e$ are defined, the goal is to find the parameter values $\boldsymbol{\beta}^s$, $\boldsymbol{\beta}^e$, $\psi^s$, $\psi^e$ and $\tau$ that maximize the DuREM likelihood function. 

Fortunately, software packages that are already suited to estimate standard Relational Event Models can be repurposed to fit a DuREM, as long as two key modifications are made. First, the statistics used in the start and end intensities must be constructed to account for the dynamic risk sets $\boldsymbol{\mathcal{R}}^s(t)$ and $\boldsymbol{\mathcal{R}}^e(t)$. After each observed event, any dyad $(i, j)$ that enters an ongoing state shifts from being at risk of starting an event to being at risk of ending it. Second, one must estimate the additional memory parameters $\psi^s$ and $\psi^e$.
A practical way to achieve this is through a grid search procedure where one specifies a two-dimensional grid of possible values for $(\psi^s, \psi^e)$, then estimates the relational event model parameters $\boldsymbol{\beta}^s$ and $\boldsymbol{\beta}^e$ for each pair, and selects the grid point that yields the maximum likelihood. 

During this grid search, for each candidate pair $(\psi^s, \psi^e)$, the software treats the duration-weighted statistics as inputs to a standard relational event model likelihood function. To make this task easier for applied researchers we have developed an open source R package \texttt{durem} that computes the statistics and estimates a DuREM model given the model specifications for the start and end rates. In addition, a parallelized grid search estimation routine is also provided to estimate the $\psi^s, \psi^e, \tau$ parameters for a given duration relational event history.

\section{Empirical applications of the DuREM}\label{sec:applications}

\subsection{Case Study I: Research team dynamics}\label{sec:case1}
In this case study, we utilize a dataset \citep{muller2018gedii} that originates from the GEDII project, which explores the impact of gender diversity on research and development teams. Specifically, it focuses on the interactions within ``team 4," a research group working at a public research centre. Over five consecutive days, sociometric badges recorded 11,607 interactions between nine team members, capturing face-to-face interactions using infrared and Bluetooth sensors. The dataset defines the duration of an event (in seconds) during which two members are in contact, excluding periods where no contact occurs within 25 seconds before or after. In addition to interaction data, the dataset provides information about team members, such as gender, age, education level, tenure, and role.

The dataset also contains relational data detailing the team's advice-seeking and social networks \citep{muller2018gedii}. The \textit{advice} matrix quantifies how the team members rated each other in a round-robin design by answering the question ``I consult this person for work related advice" (1 = never, 2 = rarely, 3 = sometimes, 4 = very often, 5 = always). Similarly, the \textit{social} matrix represents the ratings of the team members who rated each other in a round-robin design by answering the question ``I spend time socially with this person outside the lab/office" (1 = never, 2 = some times a year, 3 = some times a month, 4 = some times a week, 5 = daily) outside the workplace. 

\subsubsection{Model specification}
This dataset represents undirected relational events, as it captures interactions between pairs of individuals without specifying who initiated the contact. The events are recorded as mutual interactions detected by sensors, with the presence of both individuals in the interaction without any directionality. Hence an undirected DuREM is considered. As discussed above, DuREM consists of two components: the start rate model and the end rate model, both being undirected. Both models include a set of statistics that account for relational history, network structure, and interpersonal dynamics.

The start rate model includes various endogenous and exogenous statistics. \textit{Inertia} accounts for the influence of prior interactions, capturing the extent to which a history of events between dyads impacts the rate of initiating new interactions. \textit{Shared partners} reflects the propensity of dyads to interact based on the number of past shared partnered events between them, capturing the overlap in their social or professional network. A high value for the shared partners effect indicates greater social interconnectedness. \textit{Degree maximum} refers to the larger of the two degrees among the actors in a dyad, capturing the effect of the more connected actor on the interaction rate. \textit{Degree difference} represents the absolute difference in the number of interactions) each dyad member has within the broader network. This statistic captures interaction disparities between dyad members, with smaller degree differences imply similar levels of connectedness.

Social and advice-seeking relationships are captured using two distinct covariates for each: the average rating and the difference in ratings. For social ties, \textit{Social Tie Avg} represents the overall strength of the social bond between the dyads that may increase the likelihood of interaction. \textit{Social Tie Diff}, on the other hand, represents the difference in how each actor rates the social tie. A smaller difference implies a more reciprocal social relationship, while a larger difference indicates a disparity in perceived closeness. Similarly, for advice-seeking relationships, \textit{Advice Tie Avg} reflects the overall strength of the advice tie based on mutual ratings, with higher values indicating stronger or more trusted advisory connections. \textit{Advice Tie Diff} captures the difference in the perceived advice relationship. Lower differences suggest agreement or reciprocity in how the tie is viewed, whereas higher differences point to an imbalance in how the actors perceive the advice-seeking relationship.

%The average rating--\textit{Social tie avg} and \textit{Advice tie avg}--reflects the overall strength of the relationship between the actors of the dyad and the mutual perceptions of their bond. A higher average rating indicates a stronger or more social relationship, which may increase the tendency for interaction. In contrast, the difference in ratings--\textit{Social Tie Diff} and \textit{Advice Tie Diff}--captures the degree of asymmetry in how dyad members perceive their relationship. A lower difference indicates a balance or reciprocity in their relationship, while a higher difference reflects a disparity in perceptions. These variables allow the model to distinguish between the effects of overall relationship strength and the balance or imbalance in those relationships.

Additional statistics are included for role-related factors. For example, \textit{Same Role} identifies whether dyad members share the same functional role in the team which might influence their interaction rate due to similar or joint tasks. \textit{Same Gender} captures whether dyad members sharing the same gender have a higher or lower tendency to interact. The \textit{Professor} covariate is a dummy variable for if a professor is involved in the event. This captures the influence of hierarchical roles on interactions. %interactions involving at least one professors may be less frequent, potentially due to their limited availability. 
\textit{Difference in Tenure}, which measures the disparity in years that dyad members have worked within the research team is included to account for how professional experience differences influence interactions. %A positive coefficient for difference in tenure indicates that dyads with greater disparities in tenure are more likely to initiate interactions. 

The statistic \textit{Engaged Actor} accounts for whether either member of the dyad is concurrently involved in other events with a third actor at that time point. This could point to potential multitasking or overlapping conversations and whether actors are inclined to join ongoing events.

The end rate model includes the same set of statistics as those in the start model, but their interpretation shifts to focus on factors that affect the duration of interactions i.e rate of an event ending. For example, having more shared partners might lead to interactions ending more quickly, possibly because shared connections make it easier to coordinate. A negative effect of degree difference means that dyads with similar numbers of connections tend to stay in interactions longer, likely because they are more equally involved in the network.

Social exogenous statistics such as average social ratings and differences in social ratings, as well as advice-seeking averages and differences, are also included to understand how interpersonal dynamics influence how quickly interactions come to an end. Higher average ratings in social or advice-seeking contexts reflect stronger relationships, which may lead to interactions that last longer and are less likely to end quickly. On the other hand, larger differences in how each person views the relationship can indicate imbalance. If one person sees the relationship very differently from the other, the interaction might end sooner because it feels less mutual or stable.

Role-related and situational factors can also shape how and when interactions come to an end. The same role variable captures whether dyad members hold the same functional role within the team, which may influence how efficiently their interactions are concluded. The professor variable accounts for whether a professor is involved in the interaction, allowing the model to reflect the impact of hierarchy on interaction patterns. In this context, interactions involving professors are hypothesized to last longer because of their supervisory role. Professors may engage in more detailed discussion or provide guidance, which can naturally extend the duration of the interaction. The engaged actor variable indicates whether one of the dyad members is also involved in another interaction at the same time, capturing the possible effect of multitasking or overlapping activities. Together, these variables help explain how structural positions, roles, and situational factors affect how long interactions last.

Furthermore, it should be noted that since the observation periods per day are limited to the team's working hours, the relational events are processed to calculate time differences between consecutive interactions and group them by day. This includes removal of gaps between the last event of one day and the first event of the next, so that the interevent time between off-hours is not included in the analysis.
\renewcommand{\arraystretch}{1.4}
\begin{table}[p]
    \centering
    \fontsize{10pt}{12pt}\selectfont
    \renewcommand{\arraystretch}{1.4}
    \setlength\tabcolsep{5pt} 
    \caption{Team4 result}\label{tab:team4}
    \begin{tabular}{@{}lcccc@{}}
        \toprule
        \multicolumn{5}{c}{{DuREM}} \\
        \midrule
        \multicolumn{5}{c}{\textit{Start Rate} ($\hat{\psi}_s = 0.25 $)} \\
        \midrule
         & $\boldsymbol{\hat{\beta}}$ & se  &z-value & {p-value} \\
        \midrule
        Baseline             & -2.445 & 0.078 & -31.205 & $< 2e-16$ \\
        Inertia              & 3.122  & 0.179 & 17.425  & $< 2e-16$ \\
        Degree Maximum       & -0.787 & 0.124 & -6.329  & 2.47e-10 \\
        Degree Difference    & 0.249  & 0.118 & 2.103   & 0.0355 \\
        Shared Partners      & -0.043 & 0.138 & -0.308  & 0.758 \\
        Same Gender          & 0.025  & 0.021 & 1.176   & 0.239 \\
        Same Role            & 0.079  & 0.029 & 2.742   & 0.0061 \\
        Difference in Tenure    & 0.001  & 0.001 & 1.267   & 0.205 \\
        Advice Avg       & 0.047  & 0.013 & 3.761   & 0.00017 \\
        Advice Diff      & 0.022  & 0.011 & 2.007   & 0.045 \\
        Social Avg       & 0.018  & 0.016 & 1.118   & 0.263 \\
        Social Diff      & -0.002 & 0.016 & -0.146  & 0.884 \\
        Professor    & -1.058 & 0.040 & -26.494 & $< 2e-16$ \\
        Engaged Actor     & 0.032  & 0.0004 & 83.663 & $< 2e-16$ \\
               \midrule
        \multicolumn{5}{c}{\textit{End Rate} ($\hat{\psi}_e = -2.5$)} \\
        \midrule
        & $\boldsymbol{\hat{\beta}}$ & se & z-value & {p-value} \\
        \midrule
        Baseline             & -1.521 & 0.079 & -19.135 & $< 2e-16$ \\
        Inertia              & 4.400  & 0.566 & 7.774   & 7.62e-15 \\
        Degree Maximum       & 0.125  & 0.393 & 0.319   & 0.749 \\
        Degree Difference    & -0.256 & 0.380 & -0.674  & 0.501 \\
        Shared Partners      & -2.005 & 0.454 & -4.418  & 9.96e-06 \\
        Same Gender          & -0.011 & 0.021 & -0.530  & 0.596 \\
        Same Role            & 0.117  & 0.028 & 4.135   & 3.55e-05 \\
        Difference in Tenure    & 0.003  & 0.001 & 4.733   & 2.22e-06 \\
        Advice Avg       & 0.012  & 0.012 & 0.978   & 0.328 \\
        Advice Diff      & 0.032  & 0.011 & 2.923   & 0.0035 \\
        Social Avg       & 0.041  & 0.016 & 2.563   & 0.0104 \\
        Social Diff      & -0.045 & 0.016 & -2.853  & 0.0043 \\
        Professor    & -0.355 & 0.041 & -8.719  & $< 2e-16$ \\
        Engaged Actor             & 0.011  & 0.0005 & 23.324 & $< 2e-16$ \\
                \midrule
        \multicolumn{5}{c}{half-life ($\tau$ = 300 minutes)} \\
        \bottomrule
    \end{tabular}
\end{table}

\subsubsection{Results and Interpretation}
To estimate the parameters of the DuREM model, we conducted a grid search for jointly learning $\psi^s$, $\psi^e$, and $\tau$. The grid search explored a range of values for the parameters $\psi^s$, $\psi^e$, and $\tau$ to identify the parameters that maximizes the likelihood function of the DuREM. Specifically, the search covered the following ranges:

\[
\psi^s \in [-5, +5], \quad \psi^e \in [-5, +5], \quad \tau \in \{120, 300, 600, 1200, \varnothing \} \, \text{minutes}.
\]

\begin{figure}[t]
    \centering
    \begin{subfigure}[b]{0.48\textwidth}
        \includegraphics[width=\linewidth]{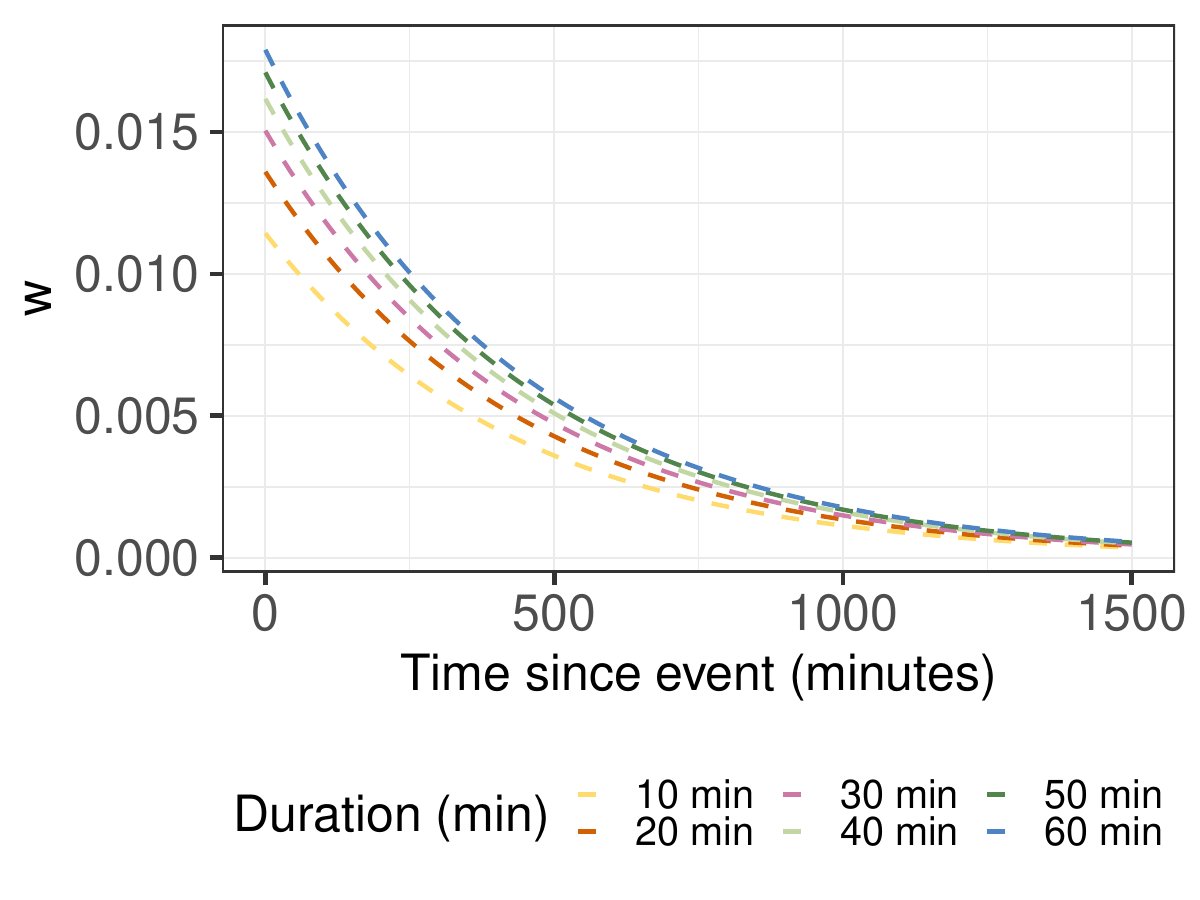}
        \caption{Event weight for $\psi^s = 0.25$, $\tau=300$}
        \label{fig:psi_comparison_s}
    \end{subfigure}
    \hfill
    \begin{subfigure}[b]{0.48\textwidth}
        \includegraphics[width=\linewidth]{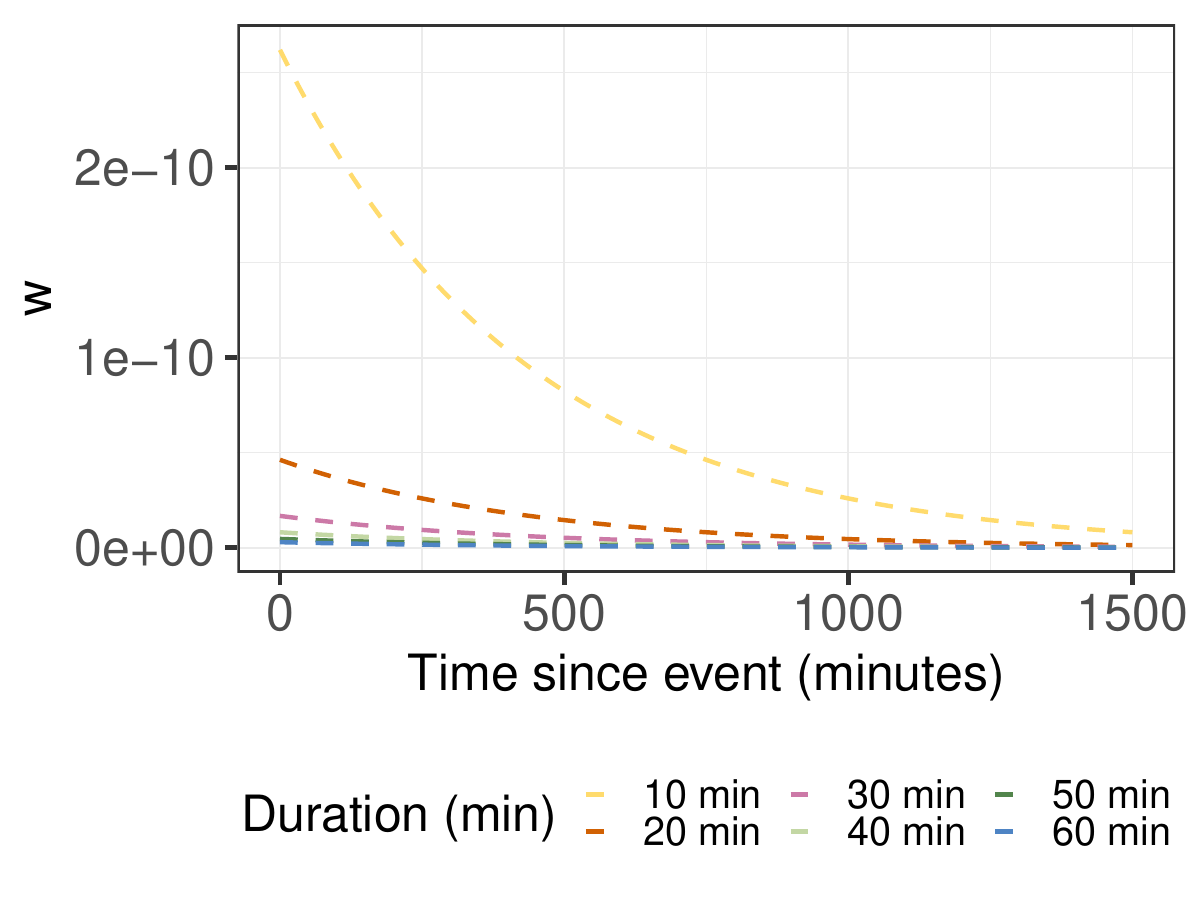}
        \caption{Event weight for $\psi^e = -2.5$, $\tau=300$}
        \label{fig:psi_comparison_e}
    \end{subfigure}
    \caption{In the first case study, the weight of past events is determined by the time since the events and their duration. The figures display event weight $w$ for the estimated $\psi^s$, $\psi^e$ and $\tau$ values.}
    \label{fig:psi_comparison}
\end{figure}

where $\tau = \varnothing$ symbolizes a DuREM with no memory effect. The analysis of the DuREM provides several important insights into how interactions are initiated and their duration within the research team. The results can be found in Table \ref{tab:team4}.

The estimated $\hat{\psi}$ parameters for start rate ($\hat{\psi}_s = 0.25$) is positive, suggesting that longer-duration events have a greater influence on the rate of starting interactions. The estimate parameter for the end rate ($\hat{\psi}_e = -2.5$) rate is negative, suggesting that shorter-duration events have a greater influence on the rate of ending interactions. This means that in the computation of endogenous statistics, such as inertia, past events with longer durations are given higher weights in the start rate model (since $\psi^s > 0$), which increases the rate at which dyads with longer past interactions initiate new events. On the other hand, in the end model, past events with longer durations are given lower weights (since $\psi^e < 0$), which decreases the rate of ending an ongoing event. Together, this suggest that dyads with longer past interactions tend to start new events at a higher rate and are less likely to end interactions quickly once they have started.

Furthermore, the half-life parameter was estimated to be 300 minutes (5 hours). So events that happened five hours ago only have half their weight left, events that happened ten hours ago only have a quarter of their weight left, etc. The difference in the magnitude of event weights  %(i.e the terms in Equation \ref{eqn:w_s} and Equation \ref{eqn:w_e}) 
for the start and end models based on the estimated {$\psi^s,\psi^e$ and $\tau$} parameters can be visualized in Figure \ref{fig:psi_comparison}. The plots illustrate the weight contribution for events with varying durations and time since event (which is relevant for the memory effect). Figure \ref{fig:psi_comparison} (\subref{fig:psi_comparison_s}) illustrates how higher duration events would contribute to a higher weight whereas Figure \ref{fig:psi_comparison} (\subref{fig:psi_comparison_e}) demonstrates higher duration events contribute to a lower weight for the end rate model.

A significant positive inertia effect in both the start and end models reveals that dyads with a high volume of mutual past interactions are more likely to initiate new interactions and tend to keep their interactions brief. Social dynamics also play an important role. Higher average advice ratings between dyads significantly increase the rate of starting interactions, while a positive effect of advice rating difference suggests that diversity in advice relations actually increases the tendency of initiating and concluding interactions. Additionally, the rate of starting an interaction increases when one of the actors is engaged in another event with a third actor, while simultaneously being engaged in another interaction makes it likely to end the current interaction between two actors more quickly.

The negative effect of degree maximum in the start model indicates that highly connected actors are less likely to initiate new dyadic interactions, potentially due to time constraints or saturation. Degree difference has a positive effect on the start rate, suggesting that dyads with larger disparities in degree are more likely to initiate interactions. Moreover, dyads with more shared partnered events tend to have longer interactions, suggesting that interactions embedded in stronger network structures are more likely to last longer. %The difference in tenure, which reflects the disparity in years worked within the research team, does not significantly influence the start rate but does affect the end rate where dyads with greater tenure disparities tend to conclude interactions more quickly.

Dyads with the same role in the research team are more likely to start interacting and also tend to conclude their interactions faster. Interactions involving professors tend to be less likely to start, but also tend to run longer. This could relate to the nature of hierarchical roles, where interactions may involve supervision or detailed discussions that require more time, professors may have limited time resources to spare (hence the lower tendency to start an interaction), but focus on interactions that require more discussion time.

The original goal of the data collection was to examine the impact of gender diversity on team interactions and dynamics. However, in this specific team, the analysis shows that gender does not have a significant effect on the rates of starting or ending interactions after controlling for other mechanisms.

These results provide a detailed view of how relational, social, and structural factors influence interaction patterns in this team. The findings show that both historical interactions and present relational attributes significantly affect how interactions are initiated and concluded. This understanding of team interaction dynamics offers a framework for analysing and improving collaborative processes in group settings, especially in environments like research teams where effective communication and efficient interaction are essential for achieving shared goals.

\subsection{Case Study II: Interpersonal Violence}\label{sec:case2}
In a second empirical application, we analyse the dynamic evolution of an interpersonal conflict in public space, drawing on the interactionist approach to the study of aggression and violence. The data set for this study includes a violent altercation between individuals, derived from CCTV footage of a public interpersonal conflict in Amsterdam, the Netherlands. The video was analysed and coded by \cite{ejbye-ernst_2022_ThirdPartiesMirror}, recording each action within the conflict with timestamps for the start and end, as well as identifiers for the initiators and recipients. Over this nine-minute episode, 220 interactions involving ten participants were recorded. The resulting relational event history of time-stamped, directed, dyadic events consisted of event durations, ranging from 0 seconds to 26.70 seconds. Seven events lacked a specific duration, such as those classified as ``hitting." These events were treated as zero-duration events. 
Furthermore, among the actors present, three were coded as ``At Work", i.e., they were employees of the establishment (such as a security guard) outside where the conflict took place.

Although many scholars attribute aggression and violence to structural factors such as racism, poverty, and sexism, or individual factors such as education, gender, and genetic predispositions, the interactionist approach emphasises the role of situational factors \citep{Felson1993,jackson-jacobs_2013_ConstructingPhysicalFights}. Within this theoretical tradition, researchers argue that once a conflict begins, its further development is predominantly shaped by endogenous situational factors \citep{Felson1984,Luckenbill1977}. Aggression and violence emerge from interactional processes, and understanding these behaviours requires consideration of their interactive nature. Therefore, explaining an individual's behaviour in a conflict situation involves accounting not only for one's preceding actions but also for the behaviours of everyone else present \citep{Felson1983}. Conflict behaviour is inherently reactive, yet this reactiveness has not been systematically studied in a manner that fully captures the complexity of preceding behaviours. Using the methods presented in this chapter, we adopt a dynamic perspective to investigate the endogenous mechanisms driving the evolution of interpersonal conflict over the course of a violent episode. Specifically, we examine how an individual’s behaviour in a conflict situation is influenced by the frequency, timing, and duration of their own prior actions and those of others present.

\subsubsection{Model specification}
In this study, we employ a Duration Relational Event Model (DuREM) to investigate how violent or aggressive acts (ie, conflict behaviours) emerge and persist over time in a public altercation. By modelling both the start and the end of these events, we obtain a more complete overview of social interaction dynamics over the observational period. The start rate model is specified as directed, while the end rate model is undirected. This approach is appropriate because each event has a clear initiator (sender) and target (receiver) for start of the conflict events. In this context, an event might be a shove, insult, threat, intervening, or other aggressive action. However, the end of the event may occur due to mutual agreement or due to interventions from other actors, and the roles for ending the event are not specified in the data.

The start rate model is parameterized as follows. \textit{Inertia} measures whether repeated events tend to occur in the same sending–receiving pair as in previous conflict events. The \textit{Reciprocity} statistic captures the tendency of actors to reciprocate previous interactions by starting a new one.  \textit{Activity Sender} (outdegree) captures how the  frequency of an individual to  have initiated  previous events drives one's tendency to start a new one. \textit{Popularity Receiver} (indegree) captures whether individuals who are frequently targeted become more likely to be targeted again. \textit{Employee} is a role-based indicator that indicates whether one member of the dyad is a staff member. \textit{Engaged Actor} indicates whether the sender or receiver is already involved in another on-going event; this is likely to lower the rate of starting a new event between them.

The model also incorporates various recency statistics \citep{} that account for the timing and reactionary events to prior interactions. \textit{Recency Send Sender} reflects whether actors who have recently started an event are more likely to start future interactions. It captures the momentum of aggression or interaction from the initiator. \textit{Recency Send Receiver} models if dyads in which the receiver has recently acted as a sender (ie, the new event is targeted at a person who recently started an event) are more prone to future interactions. These interactions could be mediatory by intervening actors or retaliation by their recent targets. \textit{Recency Receive Sender} models whether dyads in which the sender has recently acted as a receiver are more likely to initiate future interactions. This statistic captures reactionary behaviours in conflicts, where receiving an action may trigger a reciprocal or defensive response.  \textit{Recency Receive Receiver} models whether dyads in which the receiver has recently acted as a receiver are more likely to engage in future interactions. This could indicate situations in which specific individuals are repeatedly targeted. 

For the end rate model, we include statistics that are appropriate for an undirected DuREM. \textit{Inertia} captures the propensity of dyads with many past interactions to have events that end faster (or slower). \textit{Degree Maximum} captures that individuals who are more involved in the network of interactions have events that end faster. The \textit{Degree difference} captures asymmetrical relationships where one party is predominantly more involved in conflict events than the other. The \textit{Engaged Actor} statistic looks at whether the participation of an actor in multiple concurrent events influences the duration of events. A high engagement level might lead to prolonged interactions as the actor's attention is divided among several ongoing conflicts.  When an \textit{Employee}  is involved, the events might end differently compared to those involving only non-employee participants. Their trained responses or authoritative presence could either accelerate the resolution due to professional intervention or prolong the interaction if their involvement escalates the conflict. \textit{Recency Continue} captures how the recency of a dyad's last interaction affects the duration of their current event. If a dyad has interacted recently, their ongoing event is more likely to either end quickly due to the resolution of immediate tensions or persist due to unresolved issues carrying over from their last encounter.

%We also include a set of “participation shift” indicators. While participating shifts \citep{gibson_2003_ParticipationShiftsOrder} were originally developed for conversational events, more recently they have been used in the model specification of conflict behaviours \citep{gravel_2023_RivalriesReputationRetaliation,lakdawala2025bondscreatedequaldyadic}. % \textit{Retaliation (psABBA)} represents reciprocal dynamics in which a participant who was the target of a previous event responds directly to the initial actor, highlighting tit-for-tat escalation often seen in interpersonal conflict. \textit{Intervention (psABXA)} captures the involvement of a third party stepping into the interaction, interrupting the dyadic sequence to potentially mediate, de-escalate, or redirect the aggression. \textit{Displaced conflict (psABBY)} captures aggression that may be redirected from the initial target to a third actor, and \textit{Spree (psABAY) }captures repeated events by the same individual directed toward multiple targets. 
\renewcommand{\arraystretch}{1.4}

\begin{table}[p]
    \centering
    \fontsize{10pt}{12pt}\selectfont
    \renewcommand{\arraystretch}{1.4}
    \setlength\tabcolsep{5pt} 
    \caption{Street Fight Result}\label{tab:fight}
    \begin{tabular}{@{}lcccc@{}}
        \toprule
        \multicolumn{5}{c}{{DuREM}} \\
        \midrule
        \multicolumn{5}{c}{\textit{Start Rate} ($\hat{\psi}_s = 0.25$)} \\
        \midrule
         & $\boldsymbol{\hat{\beta}}$ & se  & z-value & {p-value} \\
        \midrule
        Baseline & -6.706 & 0.152 & -44.003 & $< 2e-16$ \\
        Employee & -0.581 & 0.148 & -3.929 & 8.54e-05 \\
        Engaged & -0.273 & 0.062 & -4.374 & 1.22e-05 \\
        Inertia & 2.425 & 0.254 & 9.530 & $< 2e-16$ \\
        Reciprocity & 0.268 & 0.310 & 0.863 & 3.88e-01 \\
        Activity Sender & -1.258 & 1.024 & -1.229 & 2.19e-01 \\
        Popularity Receiver & 3.713 & 0.706 & 5.259 & 1.45e-07 \\        
        Recency Send Sender & 1.467 & 0.575 & 2.550 & 1.08e-02 \\
        Recency Receive Receiver & 2.184 & 0.386 & 5.654 & 1.57e-08 \\
        Recency Receive Sender & 1.196 & 0.399 & 3.000 & 2.70e-03 \\
        Recency Send Receiver & 3.093 & 0.334 & 9.256 & $< 2e-16$ \\                
        \midrule
        \multicolumn{5}{c}{\textit{End Rate} ($\hat{\psi}_e = 2.7$)} \\
        \midrule
         & $\boldsymbol{\hat{\beta}}$ & se  & z-value & {p-value} \\
        \midrule
        Baseline & -1.587 & 0.178 & -8.889 & $< 2e-16$ \\
        Employee & -0.427 & 0.155 & -2.747 & 6.01e-03 \\
        Engaged & -0.767 & 0.056 & -13.710 & $< 2e-16$ \\
        Inertia & -5.00e-05 & 0.041 & -1.146 & 2.52e-01 \\
        Degree Maximum & 9.13e-05 & 0.021 & 4.386 & 1.16e-05 \\
        Degree Difference & -9.00e-05 & 0.025 & -3.553 & 3.81e-04 \\
        Recency Continue & 1.577 & 0.305 & 5.166 & 2.39e-07 \\        
        \bottomrule
    \end{tabular}
\end{table}

\subsubsection{Results and Interpretation}
To estimate the parameters of the DuREM model, we performed a grid search that covered the following ranges for the parameters:

\[
\psi^s \in [-5, +5], \quad \psi^e \in [-5, +5], \quad \tau \in \{30, 60, 120, 300, \varnothing \} \, \text{seconds}.
\]
The resulting MLEs can be found in Table \ref{tab:fight}.

Both duration-weight parameters ($\psi$) are estimated to be positive, suggesting that longer-duration past events have a higher ``weight'' for subsequent interactions, whether starting or ending. For the half-life parameter, we find that the model with the maximized likelihood contained no memory effect, which implies that there is no strong memory effect present in the dataset. This is appropriate as the events occur in a short observation period. 

The estimated baseline for the end rate is less negative than the estimated baseline for the start rate, indicating that events are quicker to end than they are to begin. This aligns with the observation that, in conflict situations, many interactions are brief, often consisting of short bursts of activity punctuated by pauses. These pauses do not signify the resolution of the conflict, but rather reflect a latent tension as individuals anticipate the next action, creating a cycle of intermittent engagement.

For the start rate model, the large positive and significant coefficient for Inertia fits with previous
research arguing that conflict is relational and that these behaviours happen and repeat between specific sets of people. Typically, it is not that an individual is angry and hits people at random; rather, conflict behaviour is targeted and specific to relations between specific individuals (or in some instances groups of individuals). Popularity of receiver is strongly positive, suggesting that receivers who are frequently on the receiving end remain at high risk of being targeted again. The negative effect of Activity Sender indicates that (after accounting for inertia and popularity) senders who have initiated many past events are less likely to start yet a new event at any given point in time. One possible interpretation is that very active senders might briefly “burn out” or be forced to pause after a spree of aggression, perhaps because they are being restrained or lack the immediate opportunity to strike again. The recency effects indicate that the target of a recent aggressive act tends to return the violence (recency receive sender). Instigators of a violent event are also somewhat likely to follow an initial attack with a next attack (recency send sender). Furthermore, initiators of recent events are likely to receive events (recency send receiver). These events could be mediatory by intervening actors or retaliation by their recent targets. The effect for Employee is negative. i.e., employees do not have a higher rate of conflict events, implying that the main conflict occurs between the actors that are not employees.

In the end model, a negative coefficient for Employee interactions implies that events involving employees tend to persist longer. Probably due to employees acting in mediator roles.  We also see a negative effect for Engaged actor which indicates that if someone is already involved in another ongoing event, the current conflict they are part of is more likely to last longer. Being involved in multiple confrontations at once can stretch things out, possibly due to confusion or difficulty in focusing on one situation at a time. In contrast, the positive effects for the recency effect suggest that rapidly escalating forms of conflict tend to be shorter-lived. These patterns may suggest that the intensity of such events burn out after a short period, or the intervention of third parties seeking to de-escalate the situation tends to end events faster.

\subsubsection{Sensitivity Analysis of Maximum Likelihood Estimates to Duration Parameters in DuREM}

To investigate how variations in the duration parameters $\psi_s$ and $\psi_e$ influence the maximum likelihood estimates (MLEs) of effects $\boldsymbol{\beta}$ in the DuREM framework, we analyse the trends in the MLEs under different values of these parameters. These parameters, central to the DuREM, control how event durations influence the computation of model statistics, thereby influencing the likelihood of event start and event end, and consequently, the estimated effects of endogenous and exogenous statistics.

Using the \texttt{duremstimate.grid} function in \texttt{durem} R package, we systematically explore a grid of values for $\psi_s$ and $\psi_e$ . At each point on the grid, we compute the ML estimates of $\boldsymbol{\beta}$ by fitting the DuREM to the data. This evaluates the sensitivity of the maximum likelihood estimates (MLEs) of $\boldsymbol{\beta}$ to variations in the parameters $\psi_s$ (associated with starting events) and $\psi_e$ (associated with ending events). 

\renewcommand{\arraystretch}{1}

\begin{figure}[t]
    \centering
    \includegraphics[width=0.7\linewidth]{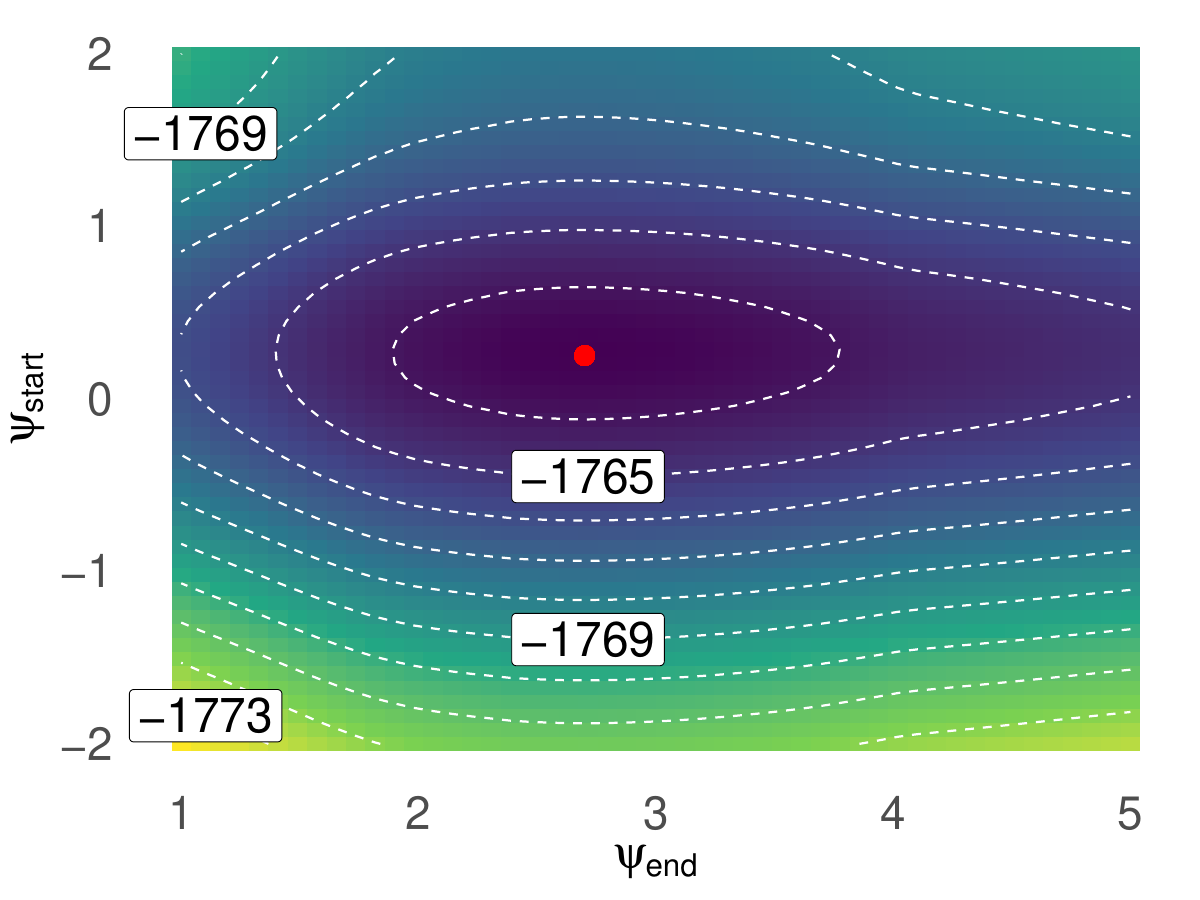}
    \caption{ Profile log-Likelihood for DuREM over a grid of $\psi$ parameters for the street fight data. The red dot indicates the MLE.}
    \label{fig:loglik}
\end{figure}

\begin{figure}[p]
     \centering
     \begin{subfigure}[b]{0.48\textwidth}
         \centering
    \includegraphics[width=\textwidth]{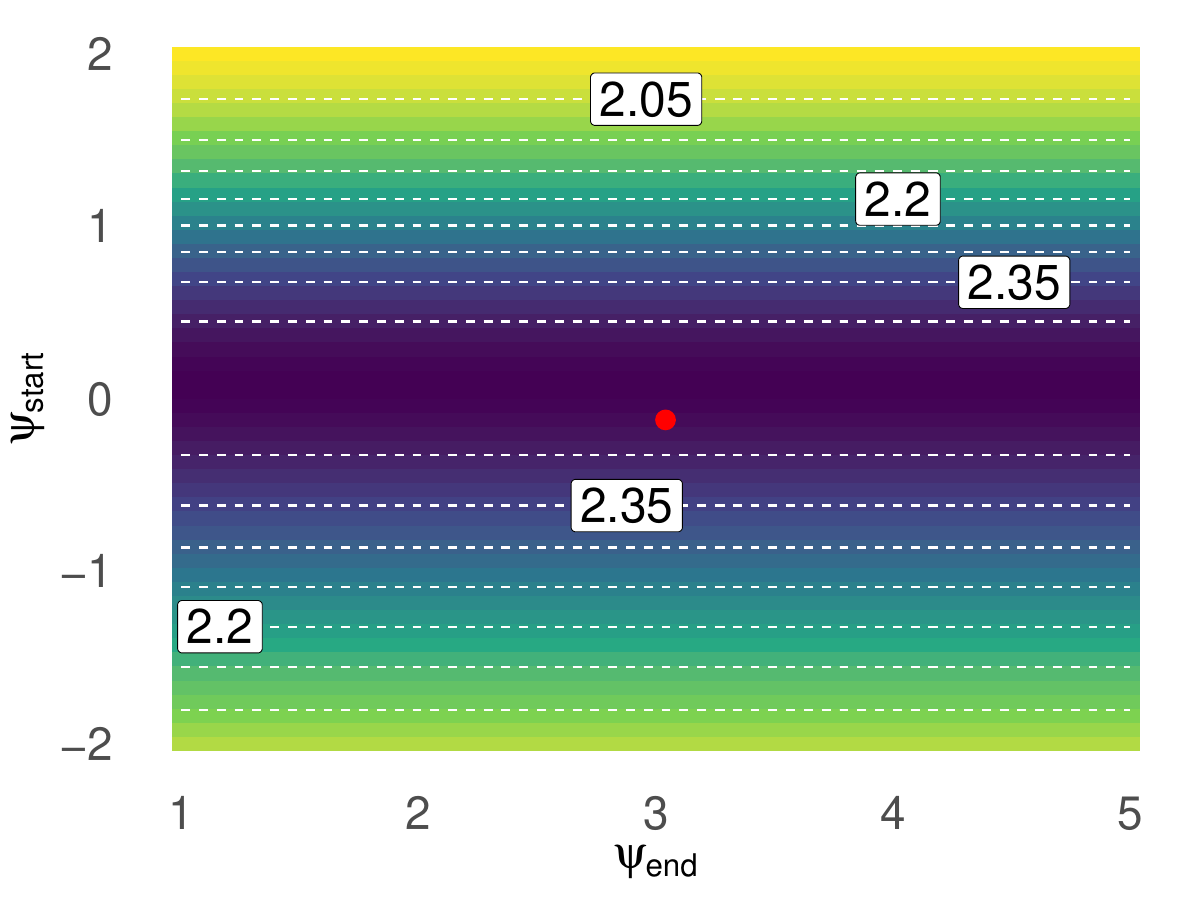}
         \caption{Inertia}
         \label{fig:gofa}
     \end{subfigure}
     %\hfill     
      %%%%%%%%%%%%%%%%%%%%%%%% VFILL %%%%%%%%%%%%%%%%%%%%%%
     \begin{subfigure}[b]{0.48\textwidth}
         \centering
         \includegraphics[width=\textwidth]{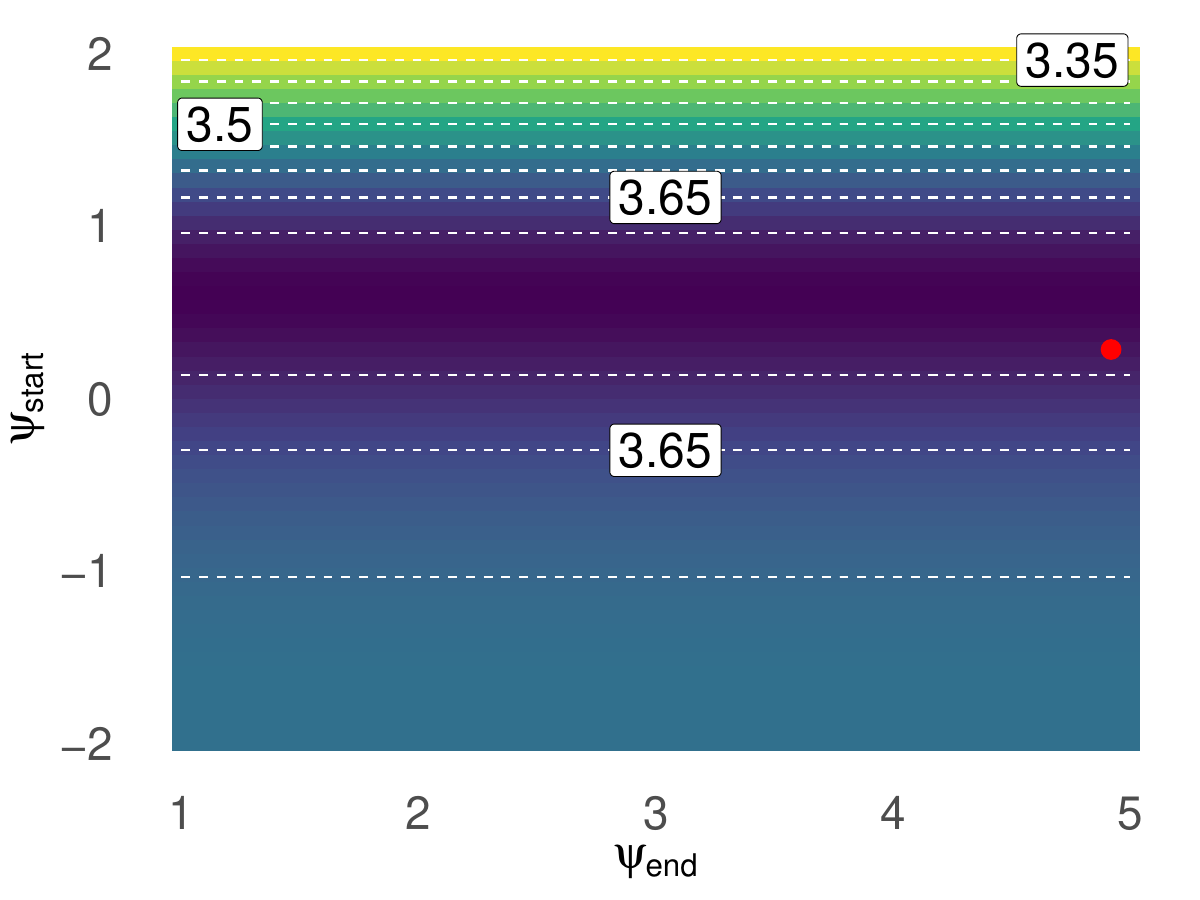}
         \caption{Popularity (in-degree) Receiver}
         \label{fig:gofc}
     \end{subfigure}
     \vfill
     \begin{subfigure}[b]{0.48\textwidth}
         \centering
         \includegraphics[width=\textwidth]{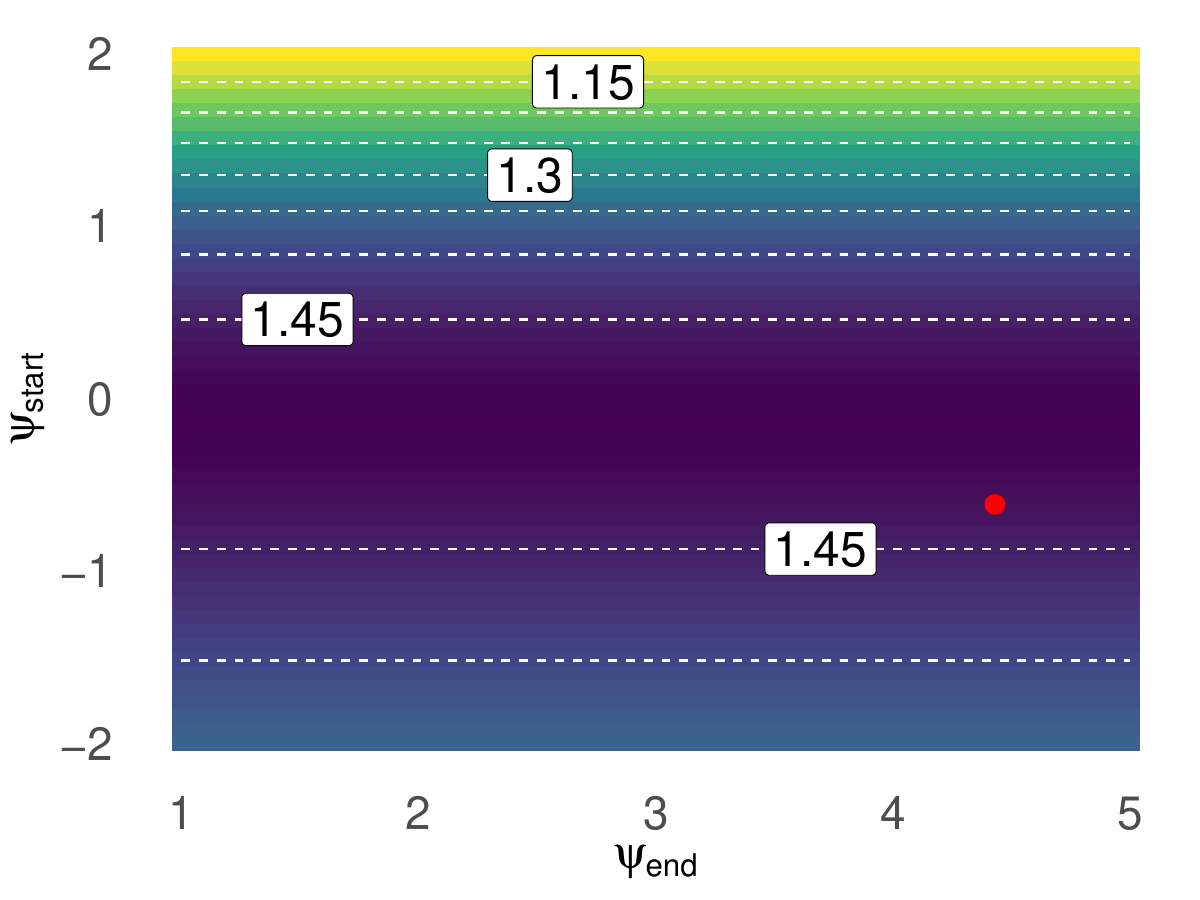}
         \caption{Recency send sender}
         \label{fig:gofd}
     \end{subfigure}
         \begin{subfigure}[b]{0.48\textwidth}
         \centering
         \includegraphics[width=\textwidth]{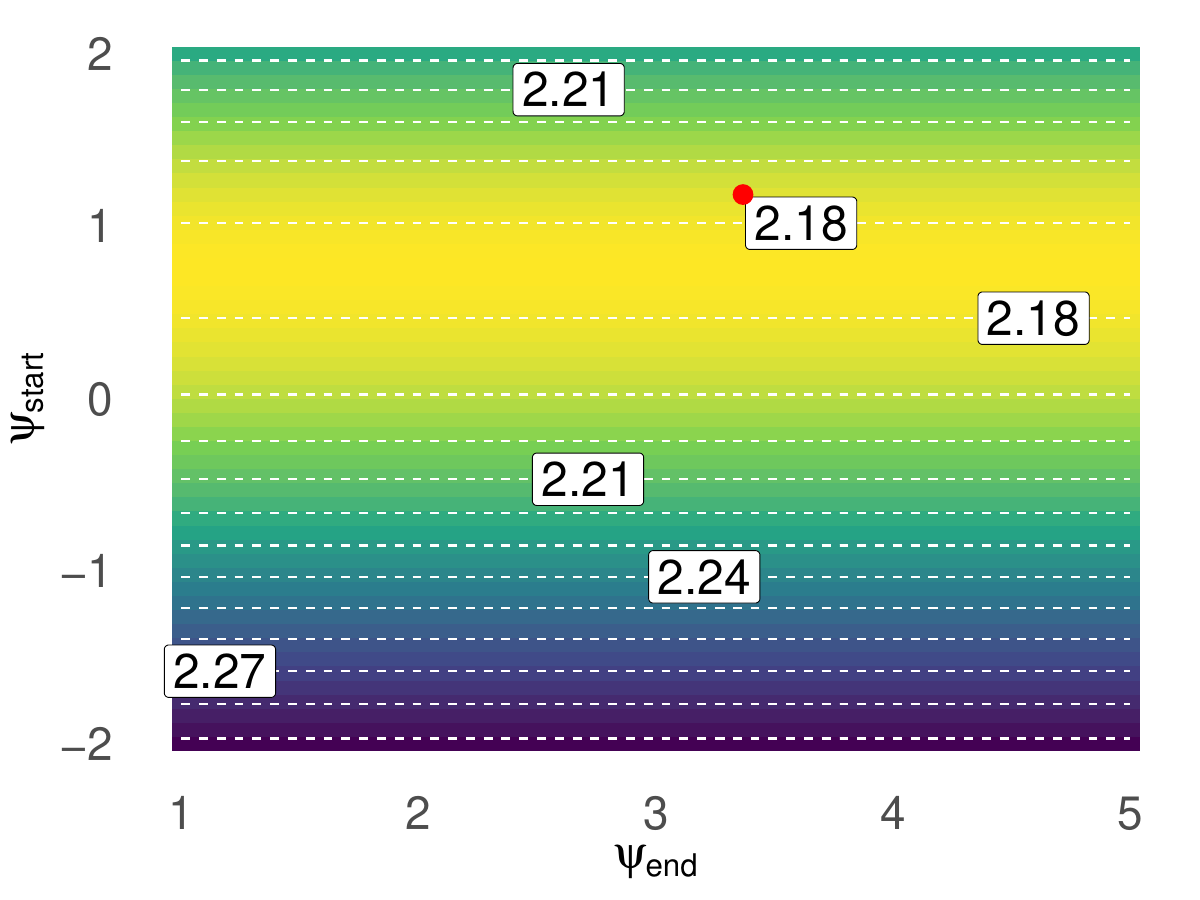}
         \caption{Recency receive receiver}
         \label{fig:gofc}
     \end{subfigure}
     \vfill
     \begin{subfigure}[b]{0.48\textwidth}
         \centering
         \includegraphics[width=\textwidth]{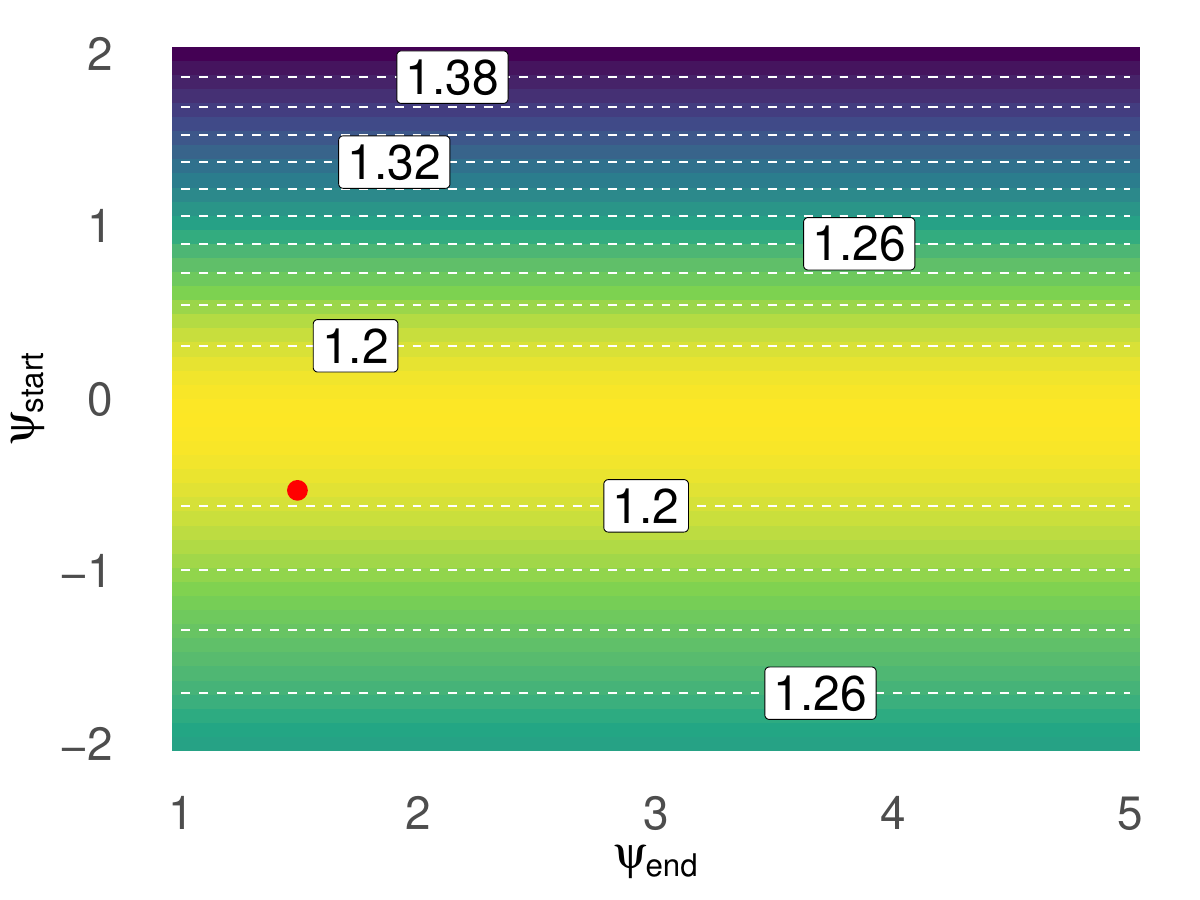}
         \caption{Recency receive sender}
     \end{subfigure}
     \begin{subfigure}[b]{0.48\textwidth}
         \centering
         \includegraphics[width=\textwidth]{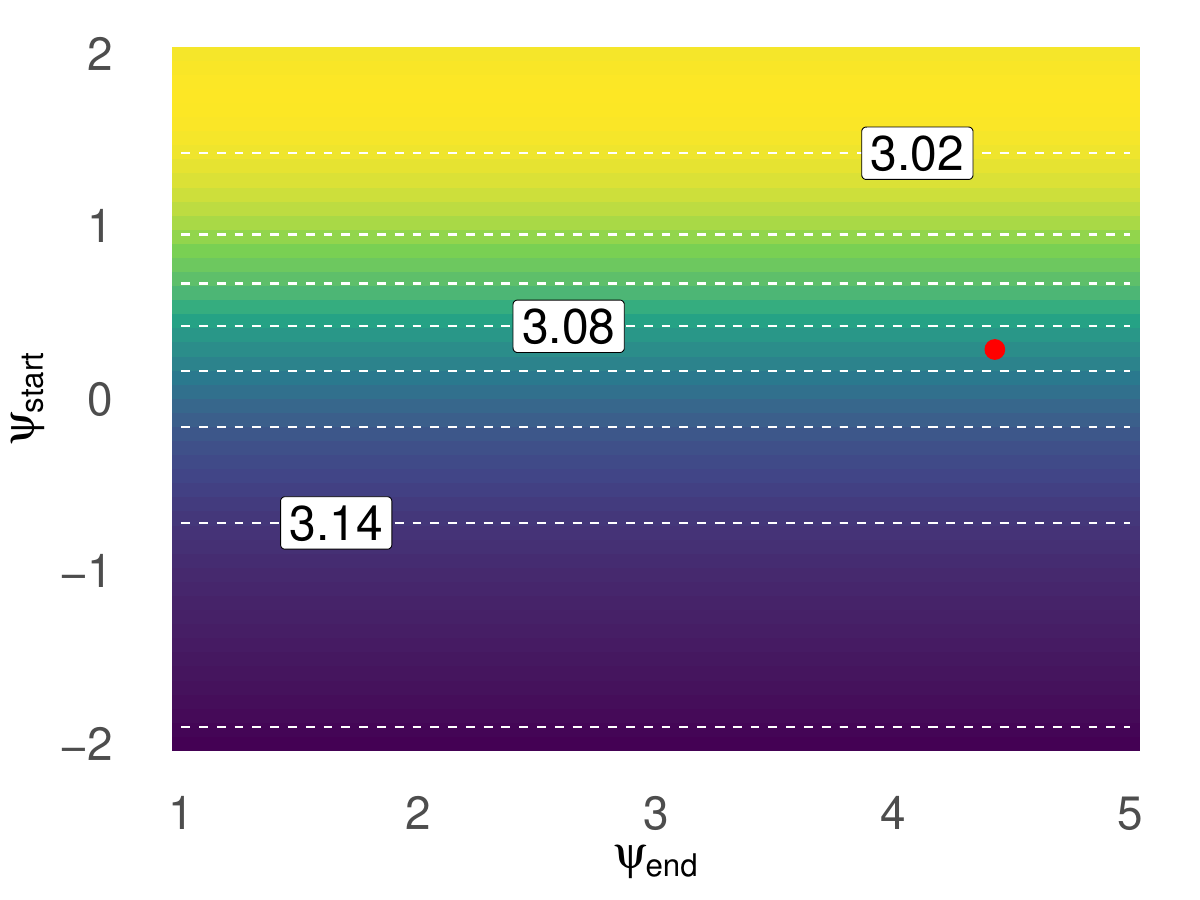}
         \caption{Recency send receiver}
     \end{subfigure}
     
     \caption{Maximum likelihood estimates of DuREM start-rate parameters ($\boldsymbol{\beta^s}$) at different values of parameters $\psi^s,\psi^e$. Numerical labels on contour lines provide specific values of $\hat{\boldsymbol{\beta^s}}$, illustrating the magnitude of changes. The red dot represents the MLEs $\hat{\boldsymbol{\beta^s}}$.The colours represent the magnitude of the contours in that region with blue indicating higher magnitudes and yellow, lower.}
        \label{fig:durem-start}
\end{figure}

\begin{figure}[p]
     \centering     
     %\hfill
     
     \begin{subfigure}[b]{0.48\textwidth}
         \centering
         \includegraphics[width=\textwidth]{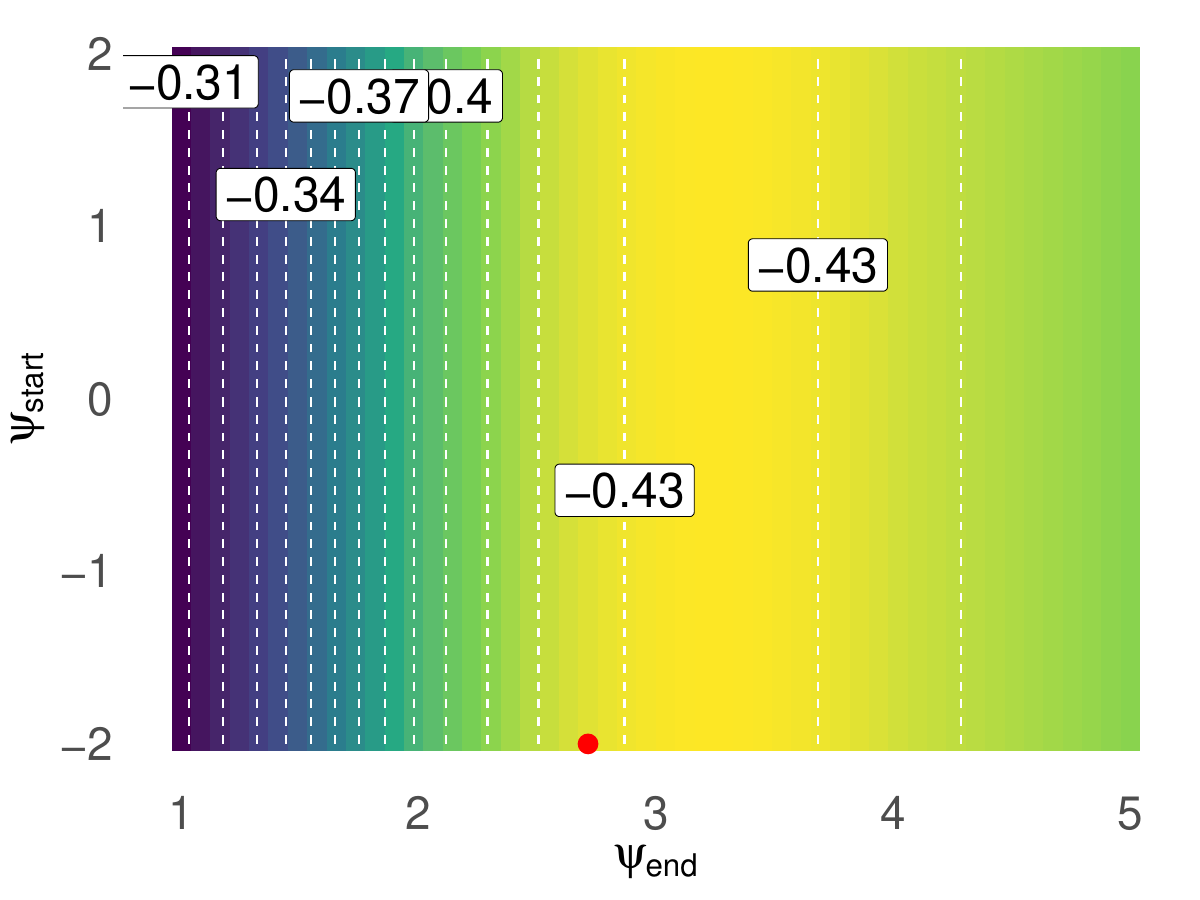}
         \caption{Employee }
         \label{fig:gofd}
     \end{subfigure}
     \begin{subfigure}[b]{0.48\textwidth}
         \centering
         \includegraphics[width=\textwidth]{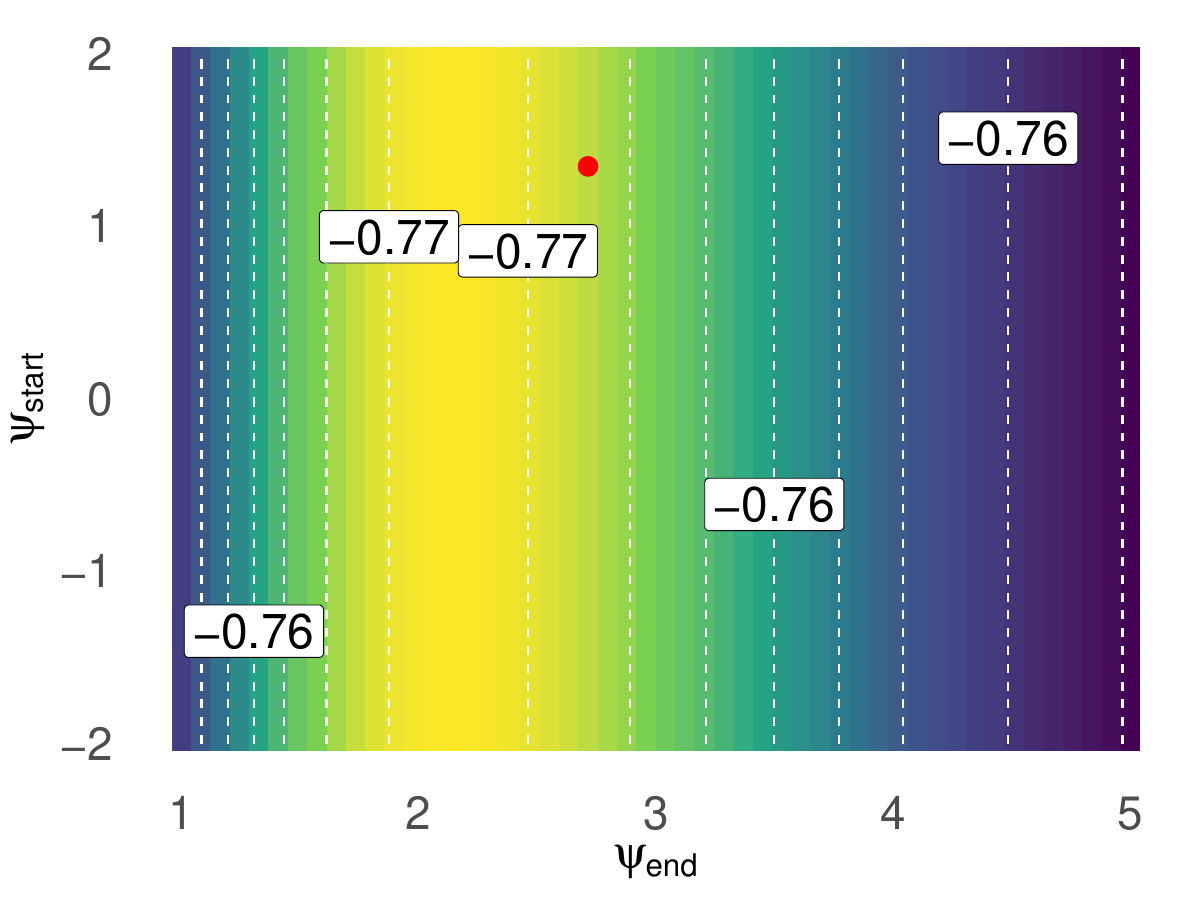}
         \caption{Engaged actor}
         \label{fig:gofd}
     \end{subfigure}
     \vfill
     \begin{subfigure}[b]{0.48\textwidth}
         \centering
         \includegraphics[width=\textwidth]{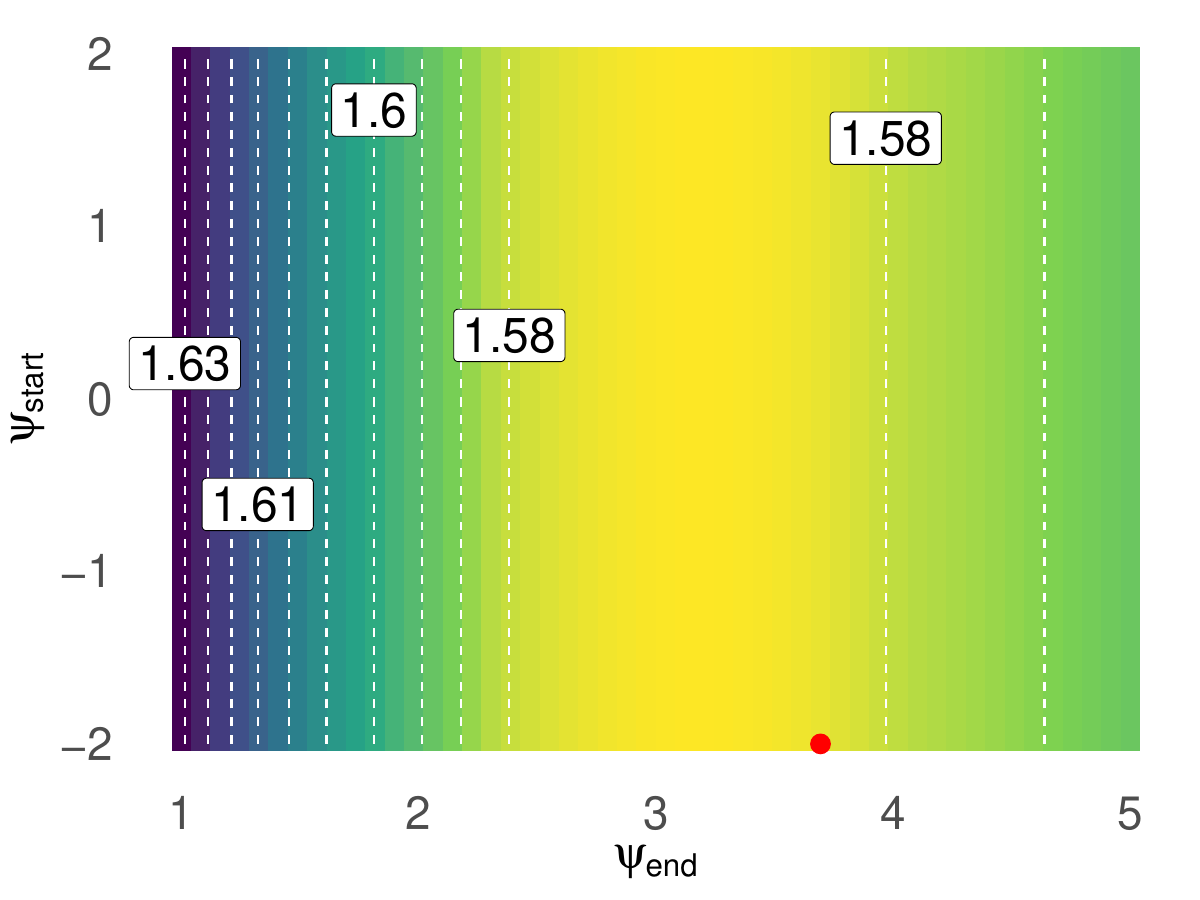}
         \caption{Recency Continue }
         \label{fig:gofd}
     \end{subfigure}
     \begin{subfigure}[b]{0.48\textwidth}
         \centering
    \includegraphics[width=\textwidth]{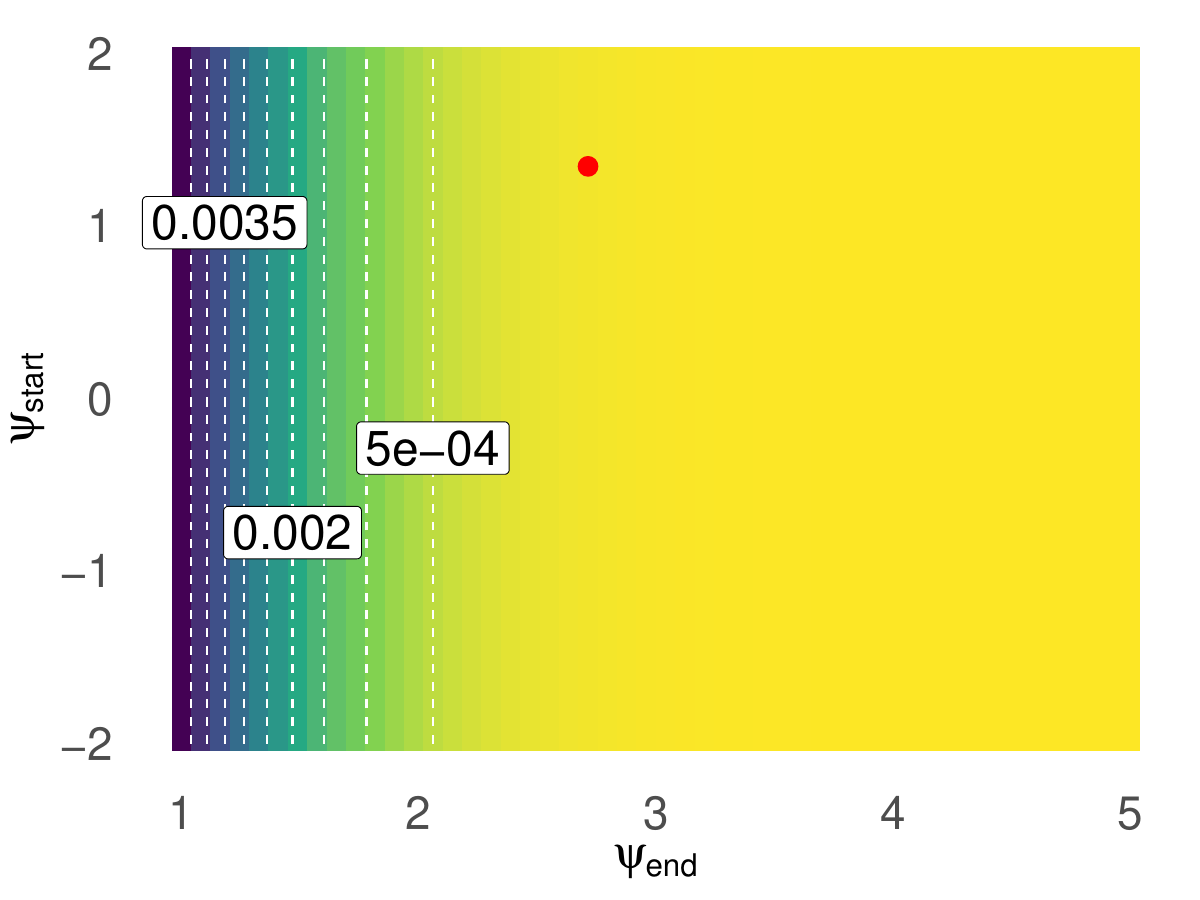}
         \caption{Degree Maximum}
         \label{fig:gofa}
     \end{subfigure}
     %\hfill
     \begin{subfigure}[b]{0.48\textwidth}
         \centering         \includegraphics[width=\textwidth]{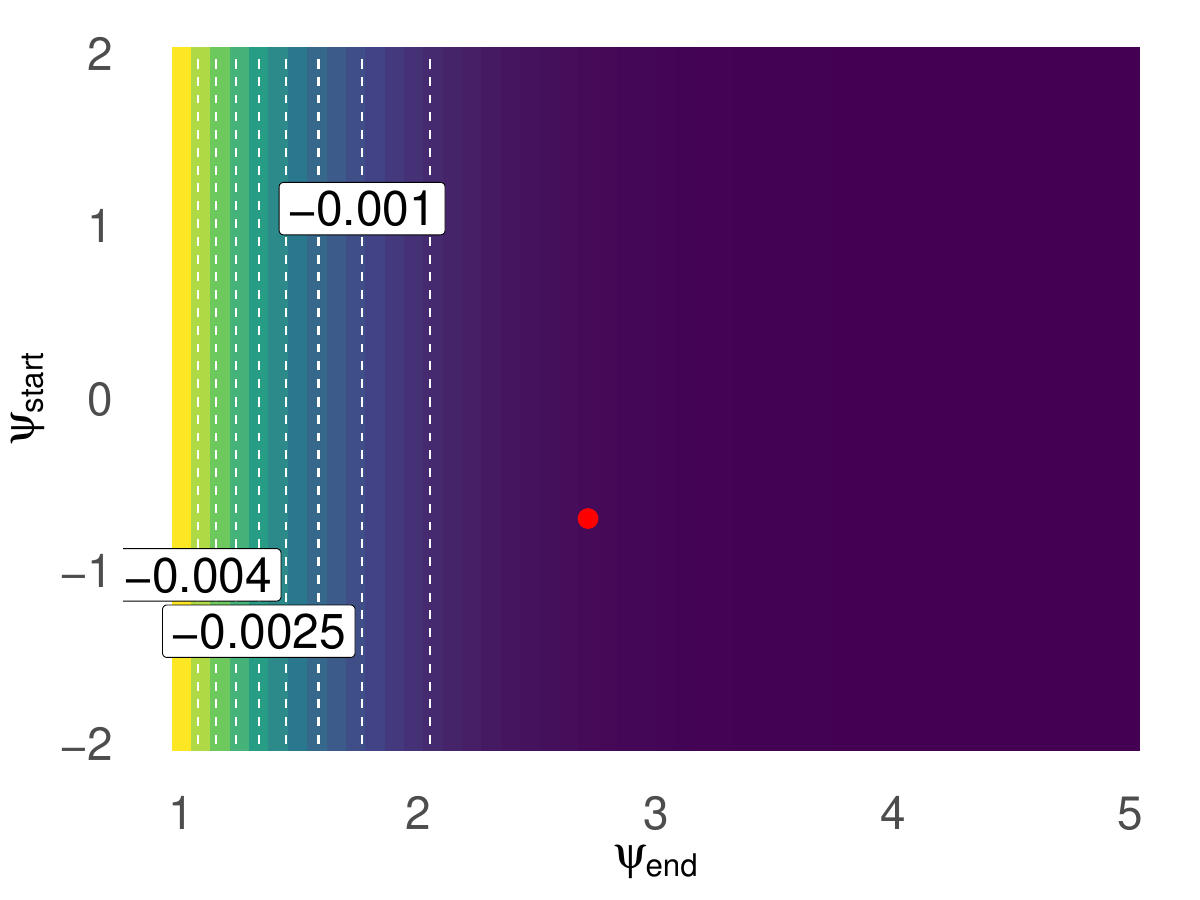}
         \caption{Degree Difference}
         \label{fig:gofc}
     \end{subfigure}
     \vfill %%%%%%%%%%%%%%%%%%%%%%%% VFILL 
     \caption{Maximum likelihood estimates of DuREM end-rate parameters ($\boldsymbol{\beta^e}$) at different values of parameters $\psi^s,\psi^e$. Numerical labels on contour lines provide specific values of $\hat{\boldsymbol{\beta^e}}$, illustrating the magnitude of changes. The red dot represents the MLEs $\hat{\boldsymbol{\beta^e}}$. The colours represent the magnitude of the contours in that region with blue indicating higher magnitudes and yellow, lower.}
        \label{fig:durem-end}
\end{figure}

The results are displayed using contour plots that illustrate the trends in the estimated effects $\hat{\boldsymbol{\beta}}$ for each statistic across the grid of $\psi_s$ and $\psi_e$.  In general, the profile log-likelihood plot in Figure \ref{fig:loglik} has a pronounced peak near the MLE, indicating that DuREM achieves a clear optimum in this parameter space. As we move away from the region the likelihood declines gradually, which indicates that the likelihood is generally well defined. 

The start rate parameters are shown in Figure \ref{fig:durem-start}, where we plot the statistically significant effects. We see that some effects exhibit relative stability across a range of $\psi_s$ and $\psi_e$, retaining their sign and changing only slightly in magnitude. However, other effects are more sensitive, with tightly clustered contours or noticeable shifts near the MLE. As either $\psi_s$ or $\psi_e$ increases or decreases, the Popularity Receiver, Recency receive-receiver, and Recency send-receiver effects preserve their sign and vary only slightly in magnitude, while Inertia, Recency send-sender, and Recency receive-sender remain consistent in sign but vary more in size in response to changing parameters $\psi$. 

The end rate parameters are shown in Figure \ref{fig:durem-end}. We observe a similar combination of stability and variability across the grid. Effects such as Employee, Engaged Actor, and Recency Continue stay consistent over a broad range of $\psi_s$ and $\psi_e$, while Degree Maximum and Degree Difference are more sensitive, with sharper changes over the duration parameters. These results show that while the DuREM allows for some flexibility in setting $\psi_s$ and $\psi_e$, selecting appropriate values is important. Still, the sensitivity of the results to the chosen values can always be evaluated afterwards and alternative models can be fitted as a result.

% table from giuseppe's thesis
% \begin{table}[h!]
%     \centering
%     \begin{tabular}{lcccccccc}
%         \toprule
%         Model & Baseline & SC & Inertia\(_{\text{start}}\) & Reciprocity\(_{\text{start}}\) & Inertia\(_{\text{end}}\) & Reciprocity\(_{\text{end}}\) & \(\psi_{\text{start}}\) & \(\psi_{\text{end}}\) \\
%         \midrule
%         start & -5.8 & 0.5 & 0.6 & 0.2 & 0.1 & 0.1 & -1.2 & 3.7 \\
%         end & \(\beta_{\text{condition}}\) & -2 & -2.9 & -4.1 & 4.0 & 0 & 1.5 & 3 \\
%         \midrule
%         \multicolumn{9}{l}{with \(\beta_{\text{condition}} = (-0.9, 0.6, 1.6, 2.5, 3.8)\)} \\
%         \bottomrule
%     \end{tabular}
%         \caption{Model parameters for network simulation}
% \end{table}

\section{Discussion}\label{sec:discuss}
%Research increasingly recognizes the role of memory decay in shaping interactions \citep[e.g.,][]{arena_2022_BayesianSemiParametricApproach, brandes_2009_NetworksEvolvingStep, perry_2013_PointProcessModelling, quintane_2013_ShortLongtermStability} and of time varying effects \citep{vieira2024fast,mulder_2019_ModelingEvolutionInteraction} but 

REMs provide a valuable statistical framework to model relational events yet, REMs assume events to be instantaneous in time and do not account for their duration. To address this shortcoming, we extended the standard REM framework for relational events with varying durations. In addition, we provide a ready to use R package \texttt{durem} to aid applied researchers in applying DuREM to various relational event data. We illustrate the applicability of the model with two case studies involving research team dynamics and interpersonal violence. The results from the case studies provide a comprehensive view of the social dynamics involving relational events and their duration and their interpretations.

Previous work \citep{meijerink-bosman_2024_RelationalEventApproach} also attempted to model the duration of relational events by extending the relational event model by considering the duration in a survival framework. There are certain key differences between the two approaches. The DuREM model decomposes the start and end of an event by modelling the interevent time between these two typed events. This approach fits naturally in the REM framework which aims to model the observed inter-event times rather than modelling the duration of each event separately. Another key difference lies in the treatment of ongoing events on the riskset. In the approach of \cite{meijerink-bosman_2024_RelationalEventApproach}, an ongoing event reduces the availability of actors in the risk set for further events. However, in practice we find that relational events between dyads often overlap, and actors can initiate new interactions while existing events are still in progress. The DuREM model addresses this limitation by allowing for such overlapping interactions, offering a more flexible and realistic representation of relational dynamics.

Moreover, \cite{hoffman_2020_ModelDynamicsFacetoface} proposed a model involving competing Poisson processes to model face-to-face interactions, similar to the approach we suggest here. In this paper, we provide a systematic approach to modelling the duration in REMs by combining the competing Poisson processes and modelling the impact of duration with the weighing scheme introduced earlier. 

While the DuREM extends the standard relational event framework by incorporating event durations, it is not without limitations. First, the assumption that actors involved in ongoing events remain eligible to be involved in new events may not always hold true. In situations where actors become unavailable by an ongoing interaction, it may be more appropriate to exclude these dyads with these actors from the risk set for new event starts. This alternative setup can naturally be incorporated by adapting the riskset for starting an event. Future research could explore such alternative formulations of the risk set where dyads containing busy or “unavailable” actors are appropriately removed.

Another limitation arises from the computational demands of the current estimation procedure. Relying on grid search for the duration-related parameters can become computationally expensive, especially in large networks or when the model incorporates numerous covariates. In such scenarios, more advanced optimization techniques (e.g., quasi-Newton methods, gradient-based algorithms, or case-controlled sampling strategies \citep{lerner_2019_ReliabilityRelationalEvent}) could be implemented to reduce computation time. Additionally, different parameterizations for the weighting scheme could be considered beyond the power function to capture the influence of duration on event rates. Another limitation is that the model assumes that event durations follow an exponential distribution, which implies a constant hazard rate between events. While this is an assumption in the standard REM, for the duration model, it may be that the probability of an event ending changes as the event progresses (e.g., longer events may be more likely to end over time). Using more flexible distributions like the Weibull may address this limitation. On the other hand, we will not be able to use the memoryless property of the exponential distribution which allows the resetting of time once an event (either a start of end) has been observed. This will make the modelling part less straightforward.

Finally, the formulation of duration itself could be refined. For instance, an alternate formulation of duration could be the proportion of time invested in specific dyadic interactions relative to an actor’s overall available interaction time in a time interval (per day, per week for instance). This approach would be particularly relevant in social contexts where actors vary in their capacity or desire for forming multiple simultaneous ties.  For instance, a project manager may divide their time across multiple team members evenly to coordinate tasks, while a specialist might focus their limited time on a single critical collaboration. Furthermore, rather than using a specific parameterized weighting function for event duration, it would also be possible to consider nonparametric approaches where the function is learned from the data. We leave these topics for future research.

\newpage
\bibliography{references}
\begin{appendices}
\section{List of statistics available in \texttt{durem}} \label{sec:stats}
In the following we provide a table of statistics available in \texttt{durem}, their formulations and the interpretation for start and end rate models.

\subsection{Endogenous DuREM Statistics for directed relational events}
{ 
\fontsize{8pt}{7pt}\selectfont
\begin{longtable}{p{2.3cm} p{4.3cm} p{4.3cm} p{4.7cm}}
\toprule
\textbf{Statistic} & \textbf{Start Rate Model \newline Interpretation} & \textbf{Duration Model \newline Interpretation} & \textbf{Formulation} \\ 
\cmidrule(lr){2-3}
& \multicolumn{2}{c}{A positive effect of this statistic implies that \dots} & {$x(i,j,t)=$} \\
\midrule
\endfirsthead
\textbf{Statistic} & \textbf{Start Rate Model \newline Interpretation} & \textbf{Duration Model \newline Interpretation} & \textbf{Formulation} \\ 
\cmidrule(lr){2-3}
& \multicolumn{2}{c}{A positive effect of this statistic implies that \dots} & {$x(i,j,t)=$} \\
\midrule
\endhead
%\midrule
\multicolumn{4}{r}{\textit{(continued on the next page)}} \\
\endfoot
\endlastfoot
inertia & dyads $(i,j)$ that have a higher intensity of previous events have a higher rate of future interaction. & dyads $(i,j)$ that have a higher intensity of previous events have a quicker tendency to end interactions. & $a(t,i,j)$ \\
reciprocity & dyads $(i,j)$ that have a higher intensity of reciprocal $(j,i)$ events have a higher rate of future interaction. & dyads that have a higher intensity of reciprocal $(j,i)$ events have a quicker tendency to end interactions. & $a(t,j,i)$ \\
indegreesender & actors that have a higher intensity of in-coming events have a higher rate of sending future interactions. & actors that have a higher intensity of in-coming events have a quicker tendency to end interactions. & $\sum \limits_k a(t,k,i)$ \\
indegreereceiver & actors $(j)$ that have a higher intensity of in-coming events have a higher rate of receiving future interactions. & actors $(j)$ that have a higher intensity of in-coming events have a quicker tendency to end interactions. & $\sum \limits_k a(t,k,j)$ \\
outdegreesender & actors $(i)$ that have a higher intensity of out-going events have a higher rate of sending future interactions. & actors $(i)$ that have a higher intensity of out-going events have a quicker tendency to end interactions. & $\sum \limits_k a(t,i,k)$ \\
outdegreereceiver & actors $(j)$ that have a higher intensity of out-going events have a higher rate of receiving future interactions. & actors $(j)$ that have a higher intensity of out-going events have a quicker tendency to end interactions. & $\sum \limits_k a(t,j,k)$ \\
totaldegreesender & actors $(i)$ with a higher total engagement (both sending and receiving) have a higher rate of future interactions. & actors $(i)$ with a higher total engagement (both sending and receiving) have a quicker tendency to end interactions. & $\sum \limits_k a(t,i,k) + a(t,k,i)$ \\
totaldegreereceiver & actors $(j)$ with a higher total engagement (both sending and receiving) have a higher rate of future interactions. & actors $(j)$ with a higher total engagement (both sending and receiving) have a quicker tendency to end interactions. & $\sum \limits_k a(t,j,k) + a(t,k,j)$ \\
itp & dyads with a higher intensity of incoming two-paths between them have a higher rate of future interaction. & dyads with a higher intensity of incoming two-paths between them have a quicker tendency to end interactions. & $\sum \limits_k \min\{a(t,j,k), a(t,k,i)\}$ \\
otp & dyads with a higher intensity of outgoing two-paths between them have a higher rate of future interaction. & dyads with a higher intensity of outgoing two-paths between them have a quicker tendency to end interactions. & $\sum \limits_k \min\{a(t,i,k), a(t,k,j)\}$ \\
isp & dyads $(i,j)$ higher intensity of in-coming shared-partnered paths have a higher rate of future interaction. & dyads $(i,j)$ that share more incoming partners have a quicker tendency to end interactions. & $\sum \limits_k \min\{a(t,k,i), a(t,k,j)\}$ \\
osp & dyads $(i,j)$ with a higher intensity of outgoing shared-partnered paths have a higher rate of future interaction. & dyads $(i,j)$ that share more outgoing partners have a quicker tendency to end interactions. & $\sum \limits_k \min\{a(t,i,k), a(t,j,k)\}$ \\
psABBA & dyads $(i,j)$ that at the previous time point had a reciprocal event $(j,i)$ have a higher rate of future interaction. & dyads $(i,j)$ that at the previous time point had a reciprocal event $(j,i)$ have a quicker tendency to end interactions. & $\begin{cases} 
1, & \text{if } \mathbb{I}(e_{t_{m-1}} = (j, i)) \\ 
0, & \text{otherwise} 
\end{cases}$ \\
psABAY & dyads $(i,j)$ where the sender remains the same as the previous sender but the receiver is a different actor have a higher rate of future interaction. & dyads $(i,j)$ where the sender remains the same as the previous sender but the receiver is a different actor have a quicker tendency to end interactions. & $\begin{cases} 
1, & \text{if } \mathbb{I}(e_{t_{m-1}} = (i, k)) \text{ \&  } j \notin \{i, k\} \\ 
0, & \text{otherwise} 
\end{cases}$ \\
psABBY & dyads $(i,j)$ where the sender is the previous receiver and the receiver is a new actor have a higher rate of future interaction. & dyads $(i,j)$ where the sender is the previous receiver and the receiver is a new actor have a quicker tendency to end interactions. & $\begin{cases} 
1, & \text{if } \mathbb{I}(e_{t_{m-1}} = (k, i)) \text{ \& } j \notin \{i, k\} \\ 
0, & \text{otherwise} 
\end{cases}$ \\
psABXA & dyads $(i,j)$ where the sender is a new actor and the receiver is the previous sender have a higher rate of future interaction. & dyads $(i,j)$ where the sender is a new actor and the receiver is the previous sender have a quicker tendency to end interactions. & $\begin{cases} 
1, & \text{if } \mathbb{I}(e_{t_{m-1}} = (k, j)) \text{ \&  } i \notin \{j, k\} \\ 
0, & \text{otherwise} 
\end{cases}$ \\
psABXB & dyads $(i,j)$ where the sender is a new actor and the receiver is the previous receiver have a higher rate of future interaction. & dyads $(i,j)$ where the sender is a new actor and the receiver is the previous receiver have a quicker tendency to end interactions. & $\begin{cases} 
1, & \text{if } \mathbb{I}(e_{t_{m-1}} = (k, i)) \text{ \&  } j \notin \{i, k\} \\ 
0, & \text{otherwise} 
\end{cases}$ \\
psABXY & dyads $(i,j)$ where both the sender and the receiver are new actors not in the previous event have a higher rate of future interaction. & dyads $(i,j)$ where both the sender and the receiver are new actors not in the previous event have a quicker tendency to end interactions. & $\begin{cases} 
1, & \text{if } \mathbb{I}(e_{t_{m-1}} = (k, l)) \text{ \&  } i,j \notin \{k, l\} \\ 
0, & \text{otherwise} 
\end{cases}$ \\
rranksend & dyads $(i,j)$ the receiver $j$ has been sent to more recently by sender $i$, with a higher rank among $i$'s past recipients, have a higher rate of future interaction. & dyads $(i,j)$ the receiver $j$ has been sent to more recently by sender $i$, with a higher rank among $i$'s past recipients, have a quicker tendency to end interactions. & $\frac{1}{r_{\text{send}}(t,i,j)}$ \\  
rrankreceive & dyads $(i,j)$ the receiver $j$ has more recently sent an event to sender $i$, with a higher rank among $i$'s past senders, have a higher rate of future interaction. & dyads $(i,j)$ the receiver $j$ has more recently sent an event to sender $i$, with a higher rank among $i$'s past senders, have a quicker tendency to end interactions. & $\frac{1}{r_{\text{receive}}(t,i,j)}$ \\  
recencysendsender & actors $(i)$ who have more recently sent an event have a higher rate of sending future interactions. & actors $(i)$ who have more recently sent an event have a quicker tendency to end interactions. & $\frac{1}{t - t_{\text{last-send}}(i) + 1}$ \\  
recencysendreceiver & dyads $(i,j)$ the receiver $j$ has more recently acted as a sender have a higher rate of future interaction. & dyads $(i,j)$ the receiver $j$ has more recently acted as a sender have a quicker tendency to end interactions. & $\frac{1}{t - t_{\text{last-send}}(j) + 1}$ \\  
recencyreceivesender & dyads $(i,j)$ the sender $i$ has more recently acted as a receiver have a higher rate of future interaction. & dyads $(i,j)$ the sender $i$ has more recently acted as a receiver have a quicker tendency to end interactions. & $\frac{1}{t - t_{\text{last-receive}}(i) + 1}$ \\  
recencyreceivereceiver & dyads $(i,j)$ the receiver $j$ has more recently acted as a receiver have a higher rate of future interaction. & dyads $(i,j)$ the receiver $j$ has more recently acted as a receiver have a quicker tendency to end interactions. & $\frac{1}{t - t_{\text{last-receive}}(j) + 1}$ \\  
recencycontinue & dyads $(i,j)$ that have interacted more recently have a higher rate of future interaction. & dyads $(i,j)$ that have interacted more recently have a quicker tendency to end interactions. & $\frac{1}{t - t_{\text{last}}(i,j) + 1}$ \\  
\bottomrule
\caption{Endogenous Statistics for Duration Relational Event Model for directed relational events. Symbols used include:  $e_t$, the event at time $t$; $\mathbb{I}(e_t = (i,j))$ indicates whether the sender and receiver pair of event at time $t$ were $(i,j)$; $w(t, i, j)$ represents the respective start or end weight of dyad $(i, j)$ at time $t$; $r_{\text{send}}(t,i,j)$ is the rank of $j$ among $i$'s most recent recipients; and $r_{\text{receive}}(t,i,j)$ is the rank of $j$ among the actors who recently sent to $i$; $a$ is defined as weighted count of past events $a(t, i, j) = \sum \limits_{o_m:(i_m, j_m) = (i, j), t_m<t} w(t, i, j)$.}
\label{tab:durem-stats}
\end{longtable}

}
\subsection{Endogenous DuREM Statistics for undirected relational events}
{
\fontsize{8pt}{7pt}\selectfont
\begin{longtable}{p{2.3cm} p{4.3cm} p{4.3cm} p{4.7cm}}
\toprule
\textbf{Statistic} & \textbf{Start Rate Model \newline Interpretation} & \textbf{Duration Model \newline Interpretation} & \textbf{Formulation} \\ 
\cmidrule(lr){2-3}
& \multicolumn{2}{c}{A positive effect of this statistic implies that \dots} & {$x(i,j,t)=$} \\
\midrule
\endfirsthead
\textbf{Statistic} & \textbf{Start Rate Model \newline Interpretation} & \textbf{Duration Model \newline Interpretation} & \textbf{Formulation} \\ 
\cmidrule(lr){2-3}
& \multicolumn{2}{c}{A positive effect of this statistic implies that \dots} & {$x(i,j,t)=$} \\
\midrule
\endhead
\midrule
\multicolumn{4}{r}{\textit{(continued on the next page)}} \\
\endfoot
\endlastfoot
inertia & dyads $(i,j)$ that have a higher intensity of previous events have a higher rate of future interaction. & dyads $(i,j)$ that have a higher intensity of previous events have a quicker tendency to end interactions. & $a(t,i,j)$ \\
totaldegreeDyad & dyads $(i,j)$ where both actors have a higher total number of past interactions have a higher rate of future interaction. & dyads $(i,j)$ where both actors have a higher total number of past interactions have a quicker tendency to end interactions. & $\sum \limits_k a(t,i,k) + \sum \limits_k a(t,j,k)$ \\
degreeMin & dyads $(i,j)$ where the lower-degree actor has a higher number of past interactions have a higher rate of future interaction. & dyads $(i,j)$ where the lower-degree actor has a higher number of past interactions have a quicker tendency to end interactions. & $\min \{ \sum \limits_k a(t,i,k), \sum \limits_k a(t,j,k) \}$ \\
degreeMax & dyads $(i,j)$ where the higher-degree actor has a higher number of past interactions have a higher rate of future interaction. & dyads $(i,j)$ where the higher-degree actor has a higher number of past interactions have a quicker tendency to end interactions. &  $\max \{ \sum \limits_k a(t,i,k), \sum \limits_k a(t,j,k) \}$ \\
degreeDiff & dyads $(i,j)$ where the difference in past interactions between the two actors is larger have a lower rate of future interaction. & dyads $(i,j)$ where the difference in past interactions between the two actors is larger have a quicker tendency to end interactions. &  $| \sum \limits_k a(t,i,k) - \sum \limits_k a(t,j,k)) |$ \\
sp & dyads $(i,j)$ that share more past shared partnered interactions have a higher rate of future interaction. & dyads $(i,j)$ that share more past shared partnered interactions have a quicker tendency to end interactions. & $\sum \limits_k \min\{a(t,i,k), a(t,j,k)\}$ \\
psABAY & dyads $(i,j)$ where one actor was involved in the previous interaction while the other actor is new have a higher rate of future interaction. & dyads $(i,j)$ where one actor was involved in the previous interaction while the other actor is new have a quicker tendency to end interactions. & $\begin{cases} 
1, & \text{if } i \text{ or } j \text{ was in } e_{t_{m-1}} \text{ and } i,j \notin e_{t_{m-1}} \\
0, & \text{otherwise} 
\end{cases}$ \\
psABAB & dyads $(i,j)$ that were part of the last interaction and reappear have a higher rate of future interaction. & dyads $(i,j)$ that were part of the last interaction and reappear have a quicker tendency to end interactions. & $\begin{cases} 
1, & \text{if } \{i, j\} = e_{t_{m-1}} \\
0, & \text{otherwise} 
\end{cases}$ \\
recencyContinue & dyads $(i,j)$ that have interacted more recently have a higher rate of future interaction. & dyads $(i,j)$ that have interacted more recently have a quicker tendency to end interactions. & $\frac{1}{t - t_{\text{last}}(i,j) + 1}$ \\  
tie & dyads $(i,j)$ with a specific exogenous dyadic attribute have a higher rate of future interaction. & dyads $(i,j)$ with a specific exogenous dyadic attribute have a quicker tendency to end interactions. & $w_{\text{tie}}(t,i,j)$ \\

\bottomrule
%\vspace{0.3cm}
\caption{Endogenous Statistics for Duration Relational Event Model for undirected relational events. Symbols used include:  $e_t$, the event at time $t$;  $\mathbb{I}(e_t = (i,j))$ indicates whether the sender and receiver pair of event at time $t$ were $(i,j)$; $w(t, i, j)$ represents the respective start or end weight of dyad $(i, j)$ at time $t$;  $a$ is defined as weighted count of past events $a(t, i, j) = \sum \limits_{o_m:(i_m, j_m) = (i, j), t_m<t} w(t, i, j)$.}
\label{tab:durem-undirected-stats}
\end{longtable}

\subsection{Exogenous DuREM Statistics}
{
\fontsize{8pt}{7pt}\selectfont
\begin{longtable}{p{2.3cm} p{4.3cm} p{4.3cm} p{4.7cm}}
\toprule
\textbf{Statistic} & \textbf{Start Rate Model \newline Interpretation} & \textbf{Duration Model \newline Interpretation} & \textbf{Formulation} \\ 
\cmidrule(lr){2-3}
& \multicolumn{2}{c}{A positive effect of this statistic implies that \dots} & {$x(i,j,t)=$} \\
\midrule
\endfirsthead
\textbf{Statistic} & \textbf{Start Rate Model \newline Interpretation} & \textbf{Duration Model \newline Interpretation} & \textbf{Formulation} \\ 
\cmidrule(lr){2-3}
& \multicolumn{2}{c}{A positive effect of this statistic implies that \dots} & {$x(i,j,t)=$} \\
\midrule
\endhead
\midrule
\multicolumn{4}{r}{\textit{(continued on the next page)}} \\
\endfoot
\endlastfoot
send & dyads $(i,j)$ where the sender $i$ actor having a higher exogenous attribute, have a higher rate of future interaction. & dyads $(i,j)$ where the sender $i$ actor having a higher exogenous attribute, have a quicker tendency to end interactions. & $c_i$ \\
receive & dyads $(i,j)$ where the receiver $j$ actor having a higher exogenous attribute, have a higher rate of future interaction. & dyads $(i,j)$ where the receive $j$ actor having a higher exogenous attribute, have a quicker tendency to end interactions. & $c_j$ \\

same & dyads $(i,j)$ where both actors share a specific exogenous attribute have a higher rate of future interaction. & dyads $(i,j)$ where both actors share a specific exogenous attribute have a quicker tendency to end interactions. & $\mathbb{I}(c_i = c_j)$ \\

difference & dyads $(i,j)$ where the difference in an exogenous actor attribute is larger have a lower rate of future interaction. & dyads $(i,j)$ where the difference in an exogenous actor attribute is larger have a quicker tendency to end interactions. & $|c_i - c_j|$ \\

average & dyads $(i,j)$ where the average value of an exogenous actor attribute is higher have a higher rate of future interaction. & dyads $(i,j)$ where the average value of an exogenous actor attribute is higher have a quicker tendency to end interactions. & $\frac{c_i + c_j}{2}$ \\

minimum & dyads $(i,j)$ where the minimum value of an exogenous actor attribute is higher have a higher rate of future interaction. & dyads $(i,j)$ where the minimum value of an exogenous actor attribute is higher have a quicker tendency to end interactions. & $\min(c_i, c_j)$ \\

maximum & dyads $(i,j)$ where the maximum value of an exogenous actor attribute is higher have a higher rate of future interaction. & dyads $(i,j)$ where the maximum value of an exogenous actor attribute is higher have a quicker tendency to end interactions. & $\max(c_i, c_j)$ \\
\bottomrule
%\vspace{0.3cm}
\caption{Endogenous Statistics for Duration Relational Event Model for undirected relational events. Symbols used include: $c_i$ which represents the value of covariate $c$ for actor $i$. }
\label{tab:durem-exo-stats}
\end{longtable}
}

}
\end{appendices}
\end{document}